\documentclass[letterpaper,10pt]{article}
\usepackage[T1]{fontenc}
\usepackage{parskip}
\usepackage[font={small}]{caption}
\usepackage[margin=2.5cm]{geometry}
\usepackage{times}
\usepackage[affil-it]{authblk}
\usepackage{listings}

\usepackage{booktabs}
\usepackage{makecell}

\date{}

\newenvironment{frontmatter}%
  {}%
  {}

\usepackage{amsmath,amssymb}
\usepackage{graphicx} 
\usepackage{subfig}
\usepackage{epsfig}
\usepackage{algorithm}

\usepackage{algpseudocode}
\usepackage{float}
\usepackage[utf8]{inputenc}
\usepackage[english]{babel}
\usepackage{amsthm}

\newtheorem{theorem}{Theorem}[section]

\newtheorem{lem}[theorem]{Lemma}

\newcommand{\nc}{\newcommand}
\nc{\mbf}{\mathbf}
\nc{\R}{\Bbb{R}}
\nc{\N}{\Bbb{N}}
\nc{\U}{\Bbb{U}}
\nc{\X}{\Bbb{X}}
\nc{\intr}[1]{\textnormal{int}#1}
\nc{\mpart}[2]{\frac{\partial #1}{\partial #2}}
\nc{\sfrac}[2]{{\textstyle\frac{#1}{#2}}}
\nc{\mprb}[2][]{\ensuremath{({\cal #2}_{#1})}}
\nc{\nser}[3]{#1_{#2},\ldots,#1_{#3}}
\nc{\nserf}[4]{#1(#2_{#3}),\ldots,#1(#2_{#4})}
\nc{\inprod}[2]{\langle#1,#2\rangle}
\nc{\ndthm}{\hspace*{\fill}$\scriptstyle{\Box}$}
\nc{\vndthm}{~\vspace{-\baselineskip}~\ndthm}
\nc{\wideq}[1][=]{\ensuremath{\ \ #1\ \ }}
\nc{\xs}{{\mathscr X}}
\nc{\us}{{\mathscr U}}
\nc{\Seq}{\Bbb{S}}
\nc{\range}[1]{\textnormal{ran}(#1)}
\nc{\domain}[1]{\textnormal{dom}(#1)}
\nc{\fdef}[3]{#1:#2,\quad #3}
\nc{\bmat}[1]{\begingroup\setlength\arraycolsep{8pt}\begin{bmatrix} #1 \end{bmatrix}\endgroup}
\nc{\cmat}[1]{\begingroup\setlength\arraycolsep{8pt} \left ( \begin{matrix} #1 \end{matrix} \right ) \endgroup}
\nc{\rank}{\textnormal{rank}}
\nc{\FF}{\mathcal{F}}
\nc{\PP}{P}
\nc{\OO}{\Omega}
\nc{\Prb}[1]{\textnormal{Pr}\left ( #1 \right )}
\nc{\myP}{\mathbf{P}}
\nc{\myA}{\mathbf{A}}
\nc{\myB}{\mathbf{B}}
\nc{\myQ}{\mathbf{Q}}
\nc{\myC}{\mathbf{C}}
\nc{\myD}{\mathbf{D}}
\nc{\myR}{\mathbf{R}}
\nc{\mySigma}{\boldsymbol{\Xi}}
\nc{\myPsi}{\boldsymbol{\Psi}}
\nc{\myPhi}{\boldsymbol{\Phi}}
\nc{\myPi}{\boldsymbol{\Sigma}}
\nc{\myBeta}{\boldsymbol{\Gamma}}
\nc{\myL}{\mathbf{L}}
\nc{\myJ}{\mathbf{J}}
\nc{\myI}{\mathbf{I}}
\nc{\myK}{\mathbf{K}}
\nc{\mys}{{s}}
\nc{\newspace}{\tilde{x}}
\nc{\statedim}{{n_x}}
\nc{\inputdim}{{n_u}}
\nc{\outputdim}{{n_y}}
\nc{\numPred}[1]{{M^\text{p}_{#1}}} 
\nc{\numFilt}[1]{{M^\text{f}_{#1}}} 
\nc{\numBack}[1]{{M^\text{b}_{#1}}} 
\nc{\numCorr}[1]{{M^\text{c}_{#1}}} 
\nc{\numSmooth}[1]{{M^\text{s}_{#1}}}
\nc{\myinte}{\hat{x}}
\nc{\ead}[0]{}
\nc{\address}[0]{}

\begin{document}

\begin{frontmatter}

\title{A New Smoothing Algorithm for \\
Jump Markov Linear Systems} 


\author{Mark P. Balenzuela\footnote{Corresponding author: Mark.Balenzuela@uon.edu.au}, Adrian G. Wills, Christopher Renton, and Brett Ninness}

\maketitle
Faculty of Engineering and Built Environment, The University of Newcastle, Callaghan, NSW 2308 Australia



\begin{abstract}                          
This paper presents a method for calculating the smoothed state distribution for Jump Markov Linear Systems.
More specifically, the paper details a novel two-filter smoother that
provides closed-form expressions for the smoothed hybrid state
distribution. This distribution can be expressed as a Gaussian mixture
with a known, but exponentially increasing, number of Gaussian
components as the time index increases. This is accompanied by
exponential growth in memory and computational requirements, which
rapidly becomes intractable.  To ameliorate this, we limit the number
of allowed mixture terms by employing a Gaussian mixture reduction strategy,
which results in a computationally tractable, but approximate smoothed distribution. The
approximation error can be balanced against computational complexity
in order to provide an accurate and practical smoothing algorithm that
compares favourably to existing state-of-the-art approaches.

\end{abstract}

\end{frontmatter}
\clearpage
\section{Introduction}
Abrupt and unexpected changes in system behaviour can often lead to
highly undesirable outcomes. For example, mechanical failure of
aircraft flight-control surfaces can have devastating consequences if
not detected and compensated for \cite{costa2006discrete}. This
particular example of change is caused by a system failure or fault,
but more generally there are many other possible causes of abrupt
change including environmental influences, modified operating
conditions, and reconfiguration of system networks. These types of
changes and their potential impact on system performance have been
observed in a wide range of applications including econometrics
\cite{kim1994dynamic}, telecommunications
\cite{logothetis1999expectation}, target tracking
\cite{mazor1998interacting}, and fault detection and isolation (FDI)
\cite{hashimoto2001sensor}, to name but a few.

Mitigating the potential impact of these abrupt changes relies on
timely and reliable detection of such events, which is the primary aim
of this paper. From a control perspective, modelling the possibility
of these events within a dynamic system structure has received
significant attention for several decades now
\cite{costa2006discrete}. System models that cater for these abrupt
changes are often afforded the epithets of either \emph{jump} or
\emph{switched} to indicate that the system can rapidly change
behaviour. Within this broad class of systems are the particular class
of interest in this paper, namely discrete-time
jump-Markov-linear-systems (JMLS), or as sometimes called,
switched-linear-dynamical-systems (SLDS)
\cite{barber2006expectation}. The primary reason for restricting our
attention to this subclass of systems is that they are relatively
simple, and yet offer enough flexibility to model the types of
real-world phenomena mentioned above.

In order to make this discussion more concrete, the JMLS class we are
concerned with in this paper can be expressed as 
\begin{subequations}
\label{eq:JMLSdef1}
\begin{align}
  x_{k+1} &= \mathbf{A}_k(z_k)x_k + \mathbf{B}_k(z_k)u_k + v_k,\\
  y_k &= \mathbf{C}_k(z_k)x_k + \mathbf{D}_k(z_k)u_k + e_k,
\end{align}
where $x_k \in \mathbb{R}^{\statedim}$ is the system state,
$y_k \in \R^\outputdim$ is the system output, $u_k \in \R^{\inputdim}$
is the system input, $z_k \in \{ 1,\dots ,m_k\}$ is a discrete random
variable that is often called the \emph{model index}, and the
noise terms $v_k$ and $e_k$ originate from the Gaussian
white noise process
\begin{align}
  \label{eq:autosam:3}
  v_k &\sim \mathcal{N}(v_k \mid {0},\mathbf{Q}_k(z_k)), \\
  e_k &\sim \mathcal{N}(e_k \mid {0},\mathbf{R}_k(z_k)).
\end{align}
\end{subequations}
The system matrices
$\{ \mathbf{A}_k, \mathbf{B}_k, \mathbf{C}_k, \mathbf{D}_k,
\mathbf{Q}_k, \mathbf{R}_k\}$ are allowed to randomly jump or switch
values for each time-index $k$ as a function of the model index
$z_k$.
The switch event is captured by allowing $z_k$ to transition to
$z_{k+1}$ stochastically with the probability of
transitioning from the $j_\text{th}$ model at time-index $k$ to the
$i_\text{th}$ model at time-index $k+1$ is expressed as
\begin{subequations}
\label{eq:JMLSdef2}
\begin{align}
\mathbb{P}(z_{k+1} = i | z_k = j) = T_k(i|j). 
\end{align}
The transition probabilities must satisfy typical mass function requirements
\begin{align}
  0 \leq T_k(i|j) \leq 1 \quad &\forall i, j,\\
  \sum_{i=1}^{m_{k+1}} T_k(i|j) = 1 \quad  &\forall j.
\end{align}
\end{subequations}
This transition probability mass function (PMF) encodes the type of
stochastic switching exhibited by the system. For example, consider a
situation where normal system operation is adequately modelled as a linear
state-space model, but where a known failure mode of the sensor
results in constant output reading around zero. This may be modelled
using $m_k=2$ systems modes, where for example
$z_k=1$ indicates normal operation and
$z_k=2$ indicates a fault (here $\myC=0$, $\myD=0$ and
$\myR>0$ model the observed zero reading). Assuming that normal operation
cannot be resumed following a fault, then the transition PMF for this
example might be modelled as (for some $0 < \epsilon < 1$)
\begin{align}
  \label{eq:autosam:11}
  T_k(1 \mid 1) &= 1-\epsilon, \quad &T_k(2 \mid 1) &= \epsilon,\\
  T_k(1 \mid 2) &= 0, \quad &T_k(2 \mid 2) &= 1.
\end{align}
In this and many other applications, it is vital to obtain accurate
knowledge of both the model index $z_k$ and the states
$x_k$ in order to reliably detect change, make accurate predictions
and take appropriate actions. At the same time, direct observation of
either $z_k$ or
$x_k$ is rarely, if ever, available in most practical
situations. Instead, these quantities must be inferred from available
noisy observations of the system; known as the state inference or
state estimation problem.

In general terms, the problem of estimating the state based on
available system measurements has received significant research
attention over several decades (see e.g. \cite{Jazwinski:1970}). Among
the many possible approaches, Bayesian estimation methods have emerged
as a strong contender, and this is the approach adopted in the
current paper. More specifically, the inference problem we are
targeting in this paper is the marginal smoothing problem; provided
with $N$ input-output data $u_{1:N} \triangleq
\{u_1,\ldots,u_N\}$ and $y_{1:N} \triangleq
\{y_1,\ldots,y_N\}$, determine the joint probability distribution
\begin{align}
  \label{eq:autosam:10}
  p(x_k, z_k \mid y_{1:N}).
\end{align}
Our particular interest in the smoothing problem is connected with
fault diagnosis \cite{balenzuela2018}, where smoothing can offer
significant improvements over filtering or prediction
estimates. Another important motivation for smoothing arises when
considering the system identification problem of estimating system
parameters based on observed data. Within the most successful methods
in this area is the so-called Expectation-Maximisation (EM) approach
that aims to provide maximum-likelihood estimates (see
e.g. \cite{SchonWN:2011}) and relies on fast and reliable computation
of expectation with respect to smoothed distributions.

It is well known that for linear systems with additive Gaussian noise,
the Kalman filter and associated linear smoothing techniques provide
full state distribution descriptions (see
e.g. \cite{kailath2000linear}). More generally, for a very general
class of nonlinear state-space systems, it is possible to employ
sequential Monte Carlo (SMC) approaches to provide estimates of the
smoothed state distribution (see e.g. \cite{Doucet11atutorial}).

Smoothing for the JMLS class falls somewhere in-between linear and
nonlinear smoothing.
%
%
On the one hand, it is tantalising that full knowledge
of the model index $z_k$ for each timestep $k$, written as $z_{1:N}$,
renders the problem as a linear time-varying state estimation problem,
for which linear smoothing techniques are directly
applicable~\cite{kailath2000linear}. On the other hand, na{\"i}ve
application of general SMC-based methods does not necessarily pay
attention to the highly structured JMLS model class, and may be highly
inefficient~\cite{whiteley2010efficient}.

The inherent difficulty in smoothing for the JMLS class may be
elucidated by previewing the closed-form expression for the smoothed
distribution $p(x_k, z_k \mid y_{1:N} )$ (see
Section~\ref{sec:exact-solution}), which can be expressed as an
indexed Gaussian mixture distribution (see
e.g. \cite{everitt2014finite}) as follows
\begin{align}
  &p(x_k, z_k | y_{1:N} ) =  \sum_{j=1}^{\numSmooth{k}} w_{k |
                          N}^j(z_k)\, \mathcal{N}\left ( x_k \mid 
                          \mu_{k | N}^j(z_k) , \myP_{k
                          | N}^j(z_k) \right ),   \label{eq:smoother_closed_form}
\end{align}
where $w_{k | N}^j(z_k)$ are non-negative mixture
weights. Importantly, the number of components $\numSmooth{k}$ is
given by (for a unimodal initial state distribution)
\begin{align}
  \label{eq:autosam:12}
  \numSmooth{k} = \prod_{k=1}^N m_k,
\end{align}
which for the case where $m_k=2$ for all $k=1,\ldots,N$ implies that
$\numSmooth{k} = 2^N$.  Therefore, the number of terms that must be
computed in the mixture \eqref{eq:smoother_closed_form} becomes
impractical, even for small data lengths $N$ and a modest number of
models $m_k$, which is well known~\cite{blom1988interacting,bergman2000markov,barber2006expectation,helmick1995fixed,kim1994dynamic}.
%

Perhaps not surprisingly then, the main approaches to solving the
smoothed state estimation problem for JMLS all employ some form of
approximation. Methods can be broadly categorised into two main areas:
1. modified linear smoothing strategies, and 2. dedicated SMC methods
that exploit the JMLS structure.  The former use linear estimation
theory while maintaining a practical number of components
$\numSmooth{k}$ in the mixture \eqref{eq:smoother_closed_form} for
each $z_{k}$.  In particular, to prevent the number of mixture components from
growing exponentially, these approaches employ standard mixture
reduction strategies that replace one mixture with another that has a
smaller number of components (see e.g. \cite{runnalls2007kullback}),
thus providing an approximate distribution.  The second main approach,
the so-called Rao-Blackwellized method, exploits the conditionally
linear model structure and describes the model index trajectories
$z_{1:N}$ using particle methods~\cite{whiteley2010efficient}, where
the number of particles are also moderated to practical levels.  For
the remainder of this paper we will focus on the first group of
methods.

Further categorisation of available methods can be made by considering
the two main approaches to state smoothing, namely, {\em
  forward-backward} smoothing, and {\em two-filter} smoothing. These
two methods stem from two different derivations of the smoothed state
distribution (see e.g. \cite{Doucet11atutorial}).

{\em Forward-backward} smoothers for JMLS include the \emph{second
  order generalized pseudo-Bayesian} (GPB2) \cite{kim1994dynamic} and
the \emph{expectation correction} (aSLDS EC) smoother
\cite{barber2012bayesian}. The GPB2 approach reduces the filtering and
smoothing distributions to a single Gaussian component for each model
index $z_k=\{1,\ldots,m_k\}$, whereas the expectation correction
smoother allows a more general Gaussian-mixture for each model
index. Both approaches use standard reduction techniques to achieve
this and both smoothers make unimodal forward prediction
approximations, and hence can make use of a Rauch--Tung--Striebel
(RTS) correction.  A detailed discussion of this approximation and its
justification can be found in
\cite{kim1994dynamic,barber2012bayesian}, but nevertheless this
approximation degrades the solution accuracy.

{\em Two-filter} formulations also employ a forward filtering stage
where standard reduction methods regulate the number of modes to
practical levels, but avoid the unimodal forward prediction approximations
required in the forward-backward smoother. However, the two-filter
approach is not one without challenges. It is also necessary to reduce
the number of modes during backward filtering step, and this reduction
requires some careful treatment
\cite{kitagawa1994two,helmick1995fixed,rahmathullah2014two,balenzuela2018}. As
detailed in \cite{rahmathullah2014two}, traditional reduction methods
do not necessarily apply to this case since the objects involved are
not distributions and may not be integrable over the state variables.  

Strategies have been proposed for avoiding this issue for a less
general class of systems \cite{kitagawa1994two}, called Gaussian
mixture models (GMMs). These suggestions include the use of additional
prior information in the backward filter, which forces integrability,
but ultimately degrades the estimate. A further suggestion in
\cite{kitagawa1994two} involves a batch calculation of the backward
filter in order to potentially avoid this problem. In a similar
manner, \cite{rahmathullah2014two} augments observations and performs
a reduction in dimension when (or if) integrable likelihoods are
formed, but also prunes unlikely model sequences based off the
probability of smoothed offspring.  The original two-filter
interacting multiple model (IMM) smoother \cite{helmick1995fixed}
suggests an alternative approximation that employs pseudo-inverses and
arbitrary values.


\textbf{{\em The contributions}} of this paper are therefore:
\begin{enumerate}
\item Provide exact closed-form expressions for the smoothed
  hybrid distribution for jump Markov linear systems. 
\item As is
  well-known~\cite{alspach1972nonlinear,rahmathullah2014two}, (1)
  involves an exponentially increasing number of mixture components,
  and as such we also provide a new backward filter likelihood
  reduction method. This method merges likelihood components in a
  manner that respects the system model and maintains zeroth, first
  and second order properties of the reduced components.
\item We present a counter-example to a commonly held conjecture
  regarding differential entropy of the mixture reduction method
  employed in this paper.
\end{enumerate}
The resulting two-filter algorithm is both accurate and
computationally tractable and compares favourably with
state-of-the-art methods.

The remainder of the paper is organised as
follows. Section~\ref{sec:problem-formulation} provides a more
detailed description of the state smoothing problem considered in this
paper. In Section~\ref{sec:exact-solution} we provide an exact
solution to the smoothing
problem. Section~\ref{sec:practical-algorithm} presents a practical
algorithm where the number of modes are moderated to manageable levels
as the forward and backward filters
progress. Section~\ref{sec:simulations} provides simulations results
that compare the new algorithm with existing approaches and Section
\ref{sec:conclusion} provides some concluding remarks.

\clearpage
\section{Problem formulation}\label{sec:problem-formulation}
Assuming that the system is described by the JMLS model in
\eqref{eq:JMLSdef1}--\eqref{eq:JMLSdef2} and provided with $N$
input-output data
\begin{align}
  \label{eq:autosam:14}
  u_{1:N} \triangleq \{u_1,\ldots,u_N\}, \qquad y_{1:N} \triangleq \{y_1,\ldots,y_N\},
\end{align}
this paper is directed towards calculating the following joint
continuous-discrete hybrid smoothed distribution
\begin{align}
  \label{eq:autosam:5}
     p(x_k,z_k | y_{1:N}).
\end{align}
For ease of exposition, we introduce a hybrid continuous-discrete
state variable $\xi_k$ defined as
\begin{align}
  \label{eq:autosam:6}
  \xi_k \triangleq \{ x_k ,z_k\},
\end{align}
and note that integration with respect to $\xi_k$ is defined as $\int
f(\xi_k) d\xi_k = \sum_{z_{k}=1}^{m_k} \int f(x_k,z_k) d x_k$.
To solve the smoothing problem \eqref{eq:autosam:5}, we employ a two-filter approach, which can be
summarised as follows. By splitting the output measurements into two
parts $y_{1:k}$ and $y_{k+1:N}$, we can express the smoothed distribution as
\begin{align}
  \label{eq:autosam:15}
  p(\xi_k | y_{1:N})  & = p(\xi_k | y_{1:k}, y_{k+1:N}).
\end{align}
Straightforward application of Bayes' rule to right-hand-side
of \eqref{eq:autosam:15} provides a smoothed state distribution
according to
\begin{align}
\label{eq:autosam:17}
 p(\xi_k | y_{1:N})  & =\frac{ p(y_{k+1:N}|\xi_k,y_{1:k})\, p(\xi_k|y_{1:k})}{p(y_{k+1:N}|y_{1:k})},
\end{align}
and by the Markov property, $p(y_{k+1:N}|\xi_k,y_{1:k}) = p(y_{k+1:N}|\xi_k)$ so that
\begin{align}
  \label{eq:autosam:1}
  p(\xi_k | y_{1:N})  & =\frac{ p(y_{k+1:N}|\xi_k)\, p(\xi_k|y_{1:k})}{p(y_{k+1:N}|y_{1:k})},
\end{align}
which is known as the two-filter smoothing formulation. The above
distribution requires the forward filter $p(\xi_k|y_{1:k})$ and the
so-called backwards filter (BF) likelihood
$p(y_{k+1:N}|\xi_k)$. Importantly, both can be defined
recursively. The forward filter recursion is well known and provided
here for completeness (it relies on Bayes' rule, the Markov property
and law of total probability)
\begin{subequations}
  \label{eq:autosam:16}
\begin{align}
  \label{eq:autosam:13}
  p(\xi_k \mid y_{1:k})  &= \frac{p(y_k \mid \xi_k)\, p(\xi_k
                             \mid y_{1:k-1})}{p(y_k \mid y_{1:k-1})},\\
  p(\xi_{k+1} \mid y_{1:k} ) &= \int p(\xi_{k+1}\mid \xi_k) \, p(\xi_k \mid y_{1:k}) d\xi_{k}.
\end{align}
\end{subequations}
The backward filter recursions, which contain all of the information
from future measurements about the hybrid state, are provided by
(again, this relies on Bayes' rule, the Markov property and the law of total probability)
\begin{subequations}
\label{eq:autosam:2}
\begin{align} 
p(y_{k:N}|\xi_{k}) &=   p(y_{k}|\xi_{k})\, p(y_{k+1:N}|\xi_k),\\
p(y_{k:N}|\xi_{k-1}) &= \int p(y_{k:N}|\xi_k)   p(\xi_k| \xi_{k-1}) \, d\xi_{k}.
\end{align}
\end{subequations}
In the following section, we provide instructions for recursively
calculating the statistics of the forward filter distribution $p(\xi_k
| y_{1:k})$ and backward filter likelihood $p(y_{k+1:N} | \xi_k)$,
before providing instructions on how these objects can be used to
generate the smoothed hybrid state distribution $p(\xi_k | y_{1:N})$.
%
\clearpage
\section{The exact solution}\label{sec:exact-solution}
Here we present a generalised set of equations for calculating the
\emph{exact} hybrid smoothed distribution for the JMLS model
class. This is achieved by employing the two-filter formulations given
by \eqref{eq:autosam:1}, \eqref{eq:autosam:16} and
\eqref{eq:autosam:2}. We progress by treating the forward filter in
Section~\ref{sec:forwards-filter} and the backward filter in
Section~\ref{sec:backw-inform-filt}. The smoother will combine the
outputs from these two filters and is detailed in
Section~\ref{sec:two-filter-smoother}. This will produce solutions
that have an unmanageable number of components, which will be attended
to in Section~\ref{sec:practical_alg} by use of an approximation. All
proofs are provided in the Appendix.



\subsection{Forwards filter}\label{sec:forwards-filter}
The forward filter and predicted distributions are provided in the
following lemma. 
\begin{lem}
\label{lem:fwdfilter45}
\begin{small}
  Under the model class \eqref{eq:JMLSdef1}--\eqref{eq:JMLSdef2} with initial prior prediction
  distribution given by
    \begin{align}
      \label{eq:MCHA6100 Notes on JMLS:7}
      p(x_1,z_1) = \sum_{i=1}^{\numPred{1}} w^i_{1|0}(z_1) 
      \mathcal{N}\left(x_1 \big| \mu^i_{1
      | 0}(z_1),  \myP^i_{1 | 0}(z_1)\right),
    \end{align}
  then the subsequent filtering and prediction distributions for $k=1,\dots,N$ are given by
  \begin{align*}
    p&(x_k,z_k | y_{1:k}) = \sum_{i=1}^\numFilt{k} w^i_{k |
                              k}(z_k) \, \mathcal{N} \left (x_k \big|  \mu^i_{k
                              | k}(z_k),  \myP^i_{k | k}(z_k)
                              \right ),\\
    p&(x_{k+1},z_{k+1} | y_{1:k}) = \nonumber\\
    &\sum_{j=1}^{\numPred{k+1}}
       w^j_{k+1 | k}(z_{k+1}) \, 
  \mathcal{N}
  \left (x_{k+1} \big|  \mu^j_{k +1| k}(z_{k+1}),  \myP^j_{k+1 |
  k}(z_{k+1}) \right ),
  \end{align*}
  respectively, where $\numFilt{k} = \numPred{k}$ and for each $i=1,\ldots,\numPred{k}$ and $z_k=1,\ldots,m_{k}$,
\begin{subequations}
\label{eq:JMLSFcorrect}
  \begin{align}
    \label{eq:fwd_fltr}
    w^i_{k | k}(z_k) &= \frac{\tilde{w}^i_{k | k}(z_k) }{\sum_{z_k =1}^{m_k}\sum_{i=1}^{\numPred{k}} \tilde{w}^i_{k | k}(z_k)},\\
    \tilde{w}^i_{k | k}(z_k) &= w^i_{k | k-1}(z_k) \cdot \mathcal{N}\left ( y_k |
                         \eta_k^i(z_k), \mySigma_{k}^i(z_k)\right ),\\
\mu^i_{k | k}(z_k) &= \mu^i_{k | k-1}(z_k) + \myK_k^i(z_k)[y_k - \eta_k^i(z_k)],\\
    \eta_k^i(z_k) &= \myC_k(z_k) \mu^i_{k | k-1}(z_k) + \myD_k(z_k)u_k,\\
    \myP^i_{k | k}(z_k) &=  \myP^i_{k | k-1}(z_k)  - \myK_k^i(z_k) \myC_k(z_k) \myP^i_{k | k-1}(z_k),\\ 
\myK_k^i(z_k) &= \myP^i_{k | k-1}(z_k) \myC^T_k(z_k) (\mySigma_{k}^i(z_k))^{-1},\\   
    \mySigma_{k}^i(z_k) &= \ \myC_k(z_k) \myP^i_{k | k-1}(z_k) \myC^T_k(z_k)+\myR_k(z_k),
  \end{align}
\end{subequations}
and $\numPred{k+1} = m_k \cdot \numFilt{k}$, then for each $i=1,\ldots,\numFilt{k}$, \\$\ell = 1,\ldots,m_k$ and \mbox{$z_{k+1}=1,\ldots,m_{k+1}$},
\begin{subequations}
\label{eq:JMLSFpredict}
  \begin{align}
    \label{eq:MCHA6100 Notes on JMLS:15}
    w^j_{k+1 | k}(z_{k+1}) &= T_k(z_{k+1} | \ell) \cdot w^i_{k
                                | k}(\ell),\\
    \mu^j_{k+1 | k}(z_{k+1}) &= \myA_k(\ell) 
                         \mu_{k | k}^i(\ell) + \myB_k(\ell) u_k ,\\
    \myP^j_{k+1 | k}(z_{k+1}) &= \myA_k(\ell) \myP^i_{k | k}(\ell)
                                \myA^T_k(\ell) + \myQ_k(\ell) ,\\
         j &= \numFilt{k} \cdot (\ell - 1) + i.
  \end{align}
\end{subequations}
\end{small}
\end{lem}
\begin{pf}
  This lemma is included for completeness and is not considered to be
  original work. A proof of this lemma is provided in
  \cite{barber2012bayesian}.
\end{pf}
Note that the formulas for the mean $\mu^j_{k+1 |
  k}(z_{k+1})$ and covariance $\myP^j_{k+1 |
  k}(z_{k+1})$ provided in Lemma~\ref{lem:fwdfilter45} do not depend
on $z_{k+1}$. This flexibility will become useful when we discuss
reduction techniques in Section~\ref{sec:practical_alg}, in which case
the mean and covariance terms may depend on $z_{k+1}$.

\subsection{Backwards information filter}\label{sec:backw-inform-filt}
In this section we detail the backwards filter. We will make extensive
use of the so-called information form in order to express the
likelihood, which can be defined as
\begin{align}
  \label{eq:MCHA6100 Notes on JMLS:18}
  \mathcal{L} \left ( x \, |\, r,\, s,\, \myL\right ) 
  &\triangleq e^{-\frac{1}{2}\left (r + 2x^T s + x^T \myL x\right )}.
\end{align}
The utility of the information formulation is that it naturally caters
for cases where the information matrix
$\mathbf{L}$ is not invertible. This is important for capturing the
sufficient statistics of backwards filter, since they are not
guaranteed to be integrable over
$x_k$ (see e.g. \cite{kitagawa1994two}), and therefore do not always
have the same form as a Gaussian mixture distribution describing a
probability density over $x_k$.
%

The equations for calculating backward information filter (BIF) likelihoods are provided by the following lemma.
\begin{lem} \begin{small} \label{lemma:backw-filt-smooth} Under the model class
  \eqref{eq:JMLSdef1}--\eqref{eq:JMLSdef2}, then
\begin{align}
  p&(y_N | x_{N},z_N) = \sum_{i=1}^{\numCorr{N}} \mathcal{L} \left ( x_{N} \, | \, \bar{r}_{N}^i(z_{N}),\, \bar{s}_{N}^i(z_{N}),\, \bar{\myL}_{N}^i(z_{N}) \right ),
\end{align}
where $\numCorr{N} = 1$ and for $z_N=1,\ldots,m_N$,
\begin{subequations}
 \label{eq:likelihood_init}
\begin{align}
  \bar{r}_{N}^1(z_{N}) &= \zeta^T_N(z_{N}) \myR^{-1}_N(z_N) \zeta_N(z_{N}) +
    \ln{|2\pi \myR_N(z_N)|}, \\
  \bar{s}_{N}^1(z_{N}) &=  \myC^T_N(z_N)\myR^{-1}_N(z_N)\zeta_N(z_{N}),\\
  \bar{\myL}_{N}^1(z_{N}) &= \myC^T_N(z_N)\myR^{-1}_N(z_N) \myC_N(i), \\
\zeta_N(z_{N}) &= \myD_N(z_N)u_N - y_N.
\end{align}
 \end{subequations}

For subsequent $k=N-1,\ldots,1$, it follows that the backwards-propagated and corrected likelihoods are given by
\begin{subequations}
\label{eq:MCHA6100 Notes on JMLS:17}
  \begin{align}
    p(y_{k+1:N}|x_{k},z_k) 
    &= \sum_{j=1}^{\numBack{k}} \mathcal{L} \left ( x_{k}\, | \,
      r_{k}^j(z_{k}),\, s_{k}^j(z_{k}), \,
      \myL_{k}^j(z_{k}) \right ),
  \end{align}
and
  \begin{align}
    p(y_{k:N}|x_{k},z_k) 
    &= \sum_{j=1}^{\numCorr{k}} {\mathcal{L}} \left ( x_{k}\, | \,
      \bar{r}_{k}^j(z_{k}),\, \bar{s}_{k}^j(z_{k}), \,
      \bar{\myL}_{k}^j(z_{k}) \right ),
  \end{align}
   \end{subequations}
respectively, where $\numBack{k} = m_{k+1} \cdot \numCorr{k+1}$ and for each \\$\ell=1,\ldots, m_{k+1}$, $i=1,\ldots, \numBack{k+1}$ and \mbox{$z_k=1,\ldots, m_k$},
\begin{subequations}
\label{label:likerecursion}
  \begin{align}
    \myL_{k}^j(z_{k})&=\myA_{k}^T(z_{k}) \myPhi_{k}^j(z_{k}) \myA_{k}(z_{k}),\\
    s^j_k(z_k) &= \myA_k^T(z_k) \bigl [ \myPhi_k^j(z_k) \myB_{k}(z_{k})u_k
           +(\myBeta^j_k(z_k))^T\bar{s}_{k+1}^i(\ell) \bigr ]  , \\ 
       r_k^j(z_k) &= \bar{r}_{k+1}^i(\ell) - \ln{|\myBeta^j_k(z_k)|} - 2 \ln{T_k(\ell | z_k)} \nonumber \\
     &\ + \begin{bmatrix}\bar{s}_{k+1}^i(\ell)  \\ \myB_{k}(z_{k})u_k\end{bmatrix}^T \begin{bmatrix}\myPsi_k^j(z_{k}) & \myBeta_k^j(z_k) \\    (\myBeta_k^j(z_k))^T & \myPhi_{k}^j(z_{k}) \end{bmatrix} \begin{bmatrix}\bar{s}_{k+1}^i(\ell) \\ \myB_{k}(z_{k})u_k\end{bmatrix},\\
    \myBeta_k^j(z_k) &= \myI- \myQ_k(z_k) \myPhi_{k}^j(z_{k}), \\
    \myPsi_k^j(z_{k}) &= \myQ_k(z_k) \myPhi_{k}^j(z_{k}) \myQ_k(z_k) - \myQ_k(z_k),\\
    \myPhi_{k}^j(z_k)&=\Bigl (\myI + \bar{\myL}_{k+1}^i(\ell) \myQ_{k}(z_{k}) \Bigr )^{-1} \bar{\myL}_{k+1}^i(\ell),\\
       j& = \numBack{k+1} \cdot (\ell - 1) + i,
 \end{align}
\end{subequations}   
where $\myI$ is the identity matrix and $\numCorr{k} =
\numBack{k}$, then for each $j=1,\ldots,
\numBack{k}$ and $z_k=1,\ldots, m_{k}$,
\begin{subequations}
\label{eq:MCHA6100 Notes on JMLS:201}
\label{label:likerecursioncorr}
  \begin{align}
    \bar{\myL}_{k}^j(z_k) &= \myL_{k}^j(z_k) + \myC_{k}^T(z_k)\myR_{k}^{-1}(z_k)\myC_{k}(z_k),\\
    \bar{s}_{k}^j(z_k) &= s_{k}^j(z_k) + \myC_{k}^T(z_k)\myR_{k}^{-1}(z_k) \zeta_{k}(z_k) ,\\
    \bar{r}_{k}^j(z_k) &= r_{k}^j(z_k) +\ln{|2\pi \myR_{k}(z_k)|} \nonumber \\
    &\qquad + \zeta_{k}^T(z_k) \myR_{k}^{-1}(z_k) \zeta_{k}(z_k) , \\
\zeta_{k}(z_k) &= \myD_{k}(z_k)u_k-y_{k}.
\end{align}
\end{subequations}
\end{small}
\end{lem}
\subsection{Two-filter smoother for JMLS}\label{sec:two-filter-smoother}
In the following lemma, we provide the equations for combining the forward filter distribution with the
backward filter likelihood to generate the smoothed distribution. 
\begin{lem} 
\begin{small}\label{lem:smoothing}
  The smoothed state distribution can be expressed for each $k=1,\ldots,N-1$ as
  \begin{align}
    \label{eq:MCHA6100 Notes on JMLS:13}
    p(x_k, z_k | y_{1:N}) &= \sum_{j=1}^{\numSmooth{k}} w_{k |
                               N}^j(z_k) \mathcal{N}\left ( x_k \mid
                                \mu_{k | N}^j(z_k) , \myP_{k
                               | N}^j(z_k) \right ),
  \end{align}
  where $\numSmooth{k}= \numBack{k} \cdot \numFilt{k}$ and for each $\ell=1,\ldots,\numBack{k}$,
  \mbox{$i=1,\ldots, \numFilt{k}$} and \mbox{$z_k=1,\ldots, m_k$},
\begin{subequations}
  \begin{align}
%
	{w}_{k | N}^j(z_k) &= \frac{\tilde{w}_{k | N}^j(z_k)}{\sum_{a=1}^{m_k} \sum_{p=1}^{m_{k|N}} \tilde{w}_{k | N}^p(a)},\\
    \tilde{w}_{k | N}^j(z_k) &= \frac{w^i_{k|k}(z_k) \sqrt{|\myP^j_{k|N}(z_k)|}}{\sqrt{|\myP^i_{k|k}(z_k)|}} e^{\frac{1}{2}\beta^j} ,\\
    \beta^j &= \left(\mu_{k|N}^j(z_k)\right)^T\left(\myP^j_{k|N}(z_k)\right)^{-1}\mu_{k|N}^j(z_k) \nonumber \\
    &\quad -  \left(\mu_{k | k}^i(z_k)\right)^T\left( \myP_{k | k}^i(z_k)\right )^{-1} \mu_{k | k}^i(z_k)  -r_k^\ell(z_k),\\
    \mu_{k | N}^j(z_k) &= \myP^j_{k|N}(z_k)\left(  \left ( \myP_{k | k}^i(z_k)\right )^{-1} \mu_{k | k}^i(z_k)-s^\ell_k(z_k) \right),\\
    \myP_{k | N}^j(z_k) &= \left ( \myL_k^\ell(z_k) + \left ( \myP_{k | k}^i(z_k)\right )^{-1}\right )^{-1},\\
    j &= \numFilt{k} \cdot (\ell - 1) + i.
  \end{align}
\end{subequations}
\end{small}
\end{lem}
The number of terms in both the forward and backward filter recursions
grows exponentially. This implies that the above solution is not
practical, except for cases where the number of observations $N$ is
small, or for example, where a fixed-lag smoother is required with
small lag-length. Otherwise, we are forced to maintain a practical
number of terms in these filters, which is discussed in the following
section.
\clearpage
\section{A practical algorithm}\label{sec:practical-algorithm}
\label{sec:practical_alg}
In this section we provide suitable approximations to reduce the
number of components in forward and backward filters. This ultimately
leads to a computationally tractable algorithm that is profiled
against existing approaches in Section~\ref{sec:simulations}. To
achieve this, we here employ a reduction strategy that chooses components to merge via the use of Kullback--Leibler
divergence~\cite{runnalls2007kullback}. This approach can be applied
straightforwardly during forward filter operation, but must be
applied carefully within backward filter operation.

\subsection{Kullback--Leibler reduction of Gaussian-mixtures}\label{sec:kullb-leibl-reduct}
Kullback--Leibler reduction (KLR) uses a Kullback--Leibler divergence
derived algorithm in order to repeatedly choose pairs of weighted Gaussian components that are approximated by a
single weighted Gaussian \cite{runnalls2007kullback}, i.e., 
\begin{small}
\begin{align} w_i\mathcal{N}(x|\mu_i,\mathbf{P}_i)+w_j\mathcal{N}(x|\mu_j,\mathbf{P}_j) \approx w_{ij} \mathcal{N}(x|\mu_{ij},\mathbf{P}_{ij}),\label{klredmain} \end{align}
\end{small}
where the statistics of the replacement weighted Gaussian are calculated using
moment-matching by implementation of
\begin{subequations}
\label{eq:KLred1}
\begin{align}
\mu_{ij}=&w_{i|ij}\mu_i+w_{j|ij}\mu_j,\\
\mathbf{P}_{ij}=&w_{i|ij}\mathbf{P}_i+w_{j|ij}\mathbf{P}_j\nonumber \\
&+w_{i|ij}w_{j|ij}(\mu_i-\mu_j)(\mu_i-\mu_j)^T,\\
w_{i|ij}=&\frac{w_i}{w_{ij}},\\
w_{j|ij}=& \frac{w_j}{w_{ij}},\\
w_{ij} =& w_i+w_j.
\end{align}
\end{subequations}
The $i$-$j$ pair of components to be merged has the lowest $B(i,j)$ value,
where
\begin{align}B(i,j) =& \frac{1}{2}\Big( w_{ij}\ln |\mathbf{P}_{ij}|-w_i\ln |\mathbf{P}_i| -w_j\ln |\mathbf{P}_j|\Big).\label{eq:KLredUB}\end{align}
This $B(i,j)$ value is an upper bound of the KL divergence on the overall Gaussian mixture (GM) from each successive approximation.
\subsubsection{Entropy and merging methods}
Mixture reduction strategies may lose information during the reduction
process and it is arguably desirable that the overall uncertainty
should not decrease as a result of this
loss of information~\cite{huber2009probabilistic}. From an information
theoretic perspective, this would require that the differential
entropy between the reduced and original mixtures to be non-negative, thus
affirming that uncertainty has not decreased as a result of
approximation. More precisely, when any distribution
$p(x)$ is approximated with
$q(x)$, the differential entropy can be calculated as
\begin{align}
\Delta h(p,q) =\int_{-\infty}^{\infty} p(x) \ln \left(p(x)\right)  -  q(x) \ln \left(q(x)\right) \, dx. 
\end{align}
Merging based approximations were originally thought to be favorable
over pruning or re-sampling methods due to a guarantee of non-negative
differential entropy values \cite{huber2009probabilistic}. This
conjecture is shown to not hold in general by way of counter example
below.

For convenience, we use a shorthand for a weighted normal function $\tilde{\mathcal{N}}=w\mathcal{N}(x|\mu,P)$.
Consider the three component GM defined by the statistics
\begin{align*}
\tilde{\mathcal{N}_1}: \{w=0.25,\mu=-0.9,P=1\}, \\
\tilde{\mathcal{N}_2}: \{w=0.25,\mu=0.9,P=1\}, \\
\tilde{\mathcal{N}_3}: \{w=0.5,\mu=0,P=0.1\}.
\end{align*}
Performing KL Reduction on this mixture will result in the following
values for $B(i,j)$
\begin{align*}
  B(1,2) = 0.1483, \quad B(1,3) = 0.3714, \quad
  B(2,3) = 0.3714,
\end{align*}
so that the $(1,2)$ pair will be selected since it has the lowest KL
bound. That is $\tilde{\mathcal{N}_1}$ and $\tilde{\mathcal{N}_2}$
will be merged into a new component $\tilde{\mathcal{N}_c}$ with 
%
statistics
\begin{align*}
\tilde{\mathcal{N}_c}: \{w=0.5,\mu=0,P=1.81\}.
\end{align*}
The resulting differential entropy is $-0.0177$, meaning the entropy of the mixture has decreased. 
This is because the merge has increased the apparent confidence about $x=0$. This
contradicts the conjecture in Theorem 4.4 of
\cite{huber2009probabilistic}. However, 
%
%
%
%
it should be noted that reducing a mixture to a single component using the KLR
merging technique guarantees a non-negative differential
entropy as claimed in \cite{huber2009probabilistic}.

\subsection{Likelihood reduction}
\label{sec:likelihood_red}
Similar to the forward filter, the backward filter requires an
approximation to be made to prevent the computational complexity
growing exponentially with each iteration.  Likelihood mixture
reduction presents additional challenges compared to probability
distributions.  This fundamental difficulty with merging components
from the backward filter stems from the possibility of the likelihood
functions having a constant value over a subspace of the state-space
\cite{rahmathullah2014two}.  Because of this, known function
approximators for density functions cannot be applied
straightforwardly.

Therefore, consider the reduction of two likelihood components
$\mathcal{L} \left ( x \, |\, r_i,\, s_i,\, \myL_i \right )$ and
$\mathcal{L} \left ( x \, |\, r_j,\, s_j,\, \myL_j \right )$ to a
single likelihood
$\mathcal{L} \left ( x \, |\, r_{ij},\, s_{ij},\, \myL_{ij}\right )$,
where $\{\myL_{ij},\mys_{ij},r_{ij}\}$ is to be determined from the
merge operation. We will restrict the merge operation to $(i,j)$ pairs
that satisfy a range-space condition that
\begin{align}
  \label{eq:MCHA6100 Notes on JMLS:50}
  \mathcal{R}(\myL_i) = \mathcal{R}(\myL_j), \qquad \mys_i, \mys_j
  \in \mathcal{R}(\myL_i),
\end{align}
where $\mathcal{R}(\mathbf{A}) \subseteq \R^m$ is the range of a
matrix $\mathbf{A} \in \R^{m \times n}$ and is given by
\begin{align}
  \label{eq:MCHA6100 Notes on JMLS:32}
  \mathcal{R}(\mathbf{A}) = \{ \mathbf{A}x | x \in \R^n \}.
\end{align}
Since the two input likelihood modes must satisfy a range-space
condition, we will further enforce that the output likelihood mode
should also satisfy the same range-space condition, otherwise the replacement
likelihood component offers new information that was not present in the original
modes. That is
\begin{align}
  \label{eq:MCHA6100 Notes on JMLS:51}
    \mathcal{R}(\myL_{ij}) = \mathcal{R}(\myL_i) = \mathcal{R}(\myL_j), \qquad \mys_{ij} \in \mathcal{R}(\myL_{ij}).
\end{align}
The information matrix $\myL_i$ is positive semi-definite and
symmetric, so that it affords a singular value decomposition (SVD)
\begin{align}
  \label{eq:MCHA6100 Notes on JMLS:52}
   \myL_i = \bmat{\mathbf{U} & \mathbf{Z}} \bmat{\boldsymbol{\Sigma}_i & \mathbf{0}\\ \mathbf{0} & \mathbf{0} }\bmat{\mathbf{U}^T \\ \mathbf{Z}^T},
\end{align}
where $\mathbf{0}$ is the zeros matrix, $\boldsymbol{\Sigma}_i \in \R^{d \times d}$ is a diagonal matrix whose diagonal
entries are the non-negative singular values, the columns of
$\mathbf{U} \in \R^{n \times d}$ provide an orthonormal basis for the range-space of $\myL_i$ and where the columns of $\mathbf{Z} \in \R^{n \times n-d}$
provide an orthonormal basis for the null-space of $\myL_i$.
Further, by assumption $\mathcal{R}(\myL_i) = \mathcal{R}(\myL_j) =
\mathcal{R}(\myL_{ij})$ and therefore
\begin{subequations}
\begin{align}
  \label{eq:MCHA6100 Notes on JMLS:42}
  \myL_j &= \mathbf{U} \boldsymbol{\Sigma}_j \mathbf{U}^T, \qquad \boldsymbol{\Sigma}_j = \boldsymbol{\Sigma}_j^T \succ 0,\\
  \myL_{ij} &= \mathbf{U} \boldsymbol{\Sigma}_{ij} \mathbf{U}^T, \qquad \boldsymbol{\Sigma}_{ij} = \boldsymbol{\Sigma}_{ij}^T \succ 0,\\
  \mys_i &= \mathbf{U} \eta_i, \qquad \text{for some} \quad \eta_j \in \R^d,\\
  \mys_j &= \mathbf{U} \eta_j, \qquad \text{for some} \quad \eta_i \in \R^d,\\
  \mys_{ij} &= \mathbf{U} \eta_{ij}, \qquad \text{for some} \quad \eta_{ij} \in \R^d.
\end{align}
\end{subequations}
Note that $\boldsymbol{\Sigma}_j$ and $\boldsymbol{\Sigma}_{ij}$ are not necessarily diagonal, and $\boldsymbol{\Sigma}_j$ can be conveniently computed via $\mathbf{U}^T \myL_j \mathbf{U}$.
Additionally, $\eta_i$ and $\eta_j$ can be computed using
$\mathbf{U}^Ts_i$, and $\mathbf{U}^Ts_j$ respectively.

It follows immediately from the fundamental theorem of linear algebra that 
\begin{align}
  \label{eq:MCHA6100 Notes on JMLS:40}
  x \in \mathcal{R}(\myL_i) \,  \iff \,  x = \mathbf{U} \newspace, \quad \text{for some} \quad \newspace \in \R^d.
\end{align}
Therefore, the following equalities hold (by substitution) for any
$\newspace \in \R^d$
\begin{subequations}
\begin{align}
  \label{eq:1}
  \mathcal{L}(x| r_i, \mys_i, \myL_i) &= \mathcal{L}(\newspace| r_i, \eta_i,
  \boldsymbol{\Sigma}_i),\\
  \mathcal{L}(x| r_j, \mys_j, \myL_j) &= \mathcal{L}(\newspace| r_j, \eta_j,
  \boldsymbol{\Sigma}_j),\\
  \mathcal{L}(x| r_{ij}, \mys_{ij}, \myL_{ij}) &= \mathcal{L}(\newspace| r_{ij}, \eta_{ij},
  \boldsymbol{\Sigma}_{ij}).
\end{align}
\end{subequations}
Since $\boldsymbol{\Sigma}_i, \boldsymbol{\Sigma}_j$ and $\boldsymbol{\Sigma}_{ij}$ are all full rank, then each
exponential term can be expressed as a scaled multivariate Normal
distribution according to
\begin{align}
  \label{eq:2}
  \mathcal{L}(\newspace | r, \eta,  \boldsymbol{\Sigma}) &= \alpha \ \mathcal{N} \left(\newspace\ |\boldsymbol{\Sigma}^{-1} \eta , \boldsymbol{\Sigma}^{-1} \right),
\end{align}
where 
\begin{align}
  \label{eq:MCHA6100 Notes on JMLS:44}
  \alpha = e^{-\frac{1}{2}\left(r - \eta^T\boldsymbol{\Sigma}^{-1}\eta -\ln|2
             \pi \boldsymbol{\Sigma}^{-1}|\right)} .
\end{align}
Importantly, we can then employ the standard KL reduction approach
from Section~\ref{sec:kullb-leibl-reduct} to provide
\begin{subequations}
\label{eq:5}
\begin{align}
  r_{ij}      &= \eta_{ij}^T\boldsymbol{\Sigma}_{ij}^{-1}\eta_{ij} - 2 \ln \left ( \alpha_i + \alpha_j          \right ) + \ln|2\pi \boldsymbol{\Sigma}_{ij}^{-1}|,\\
  \eta_{ij}   &= \boldsymbol{\Sigma}_{ij} \left ( v_i \mu_i + v_j \mu_j\right ) ,\\
  \boldsymbol{\Sigma}_{ij}^{-1} &= v_i \boldsymbol{\Sigma}_i^{-1} + v_j\boldsymbol{\Sigma}_j^{-1} + v_i v_j \left
                ( \mu_i - \mu_j\right ) \left ( \mu_i - \mu_j \right )^T,\\
  \mu_i  &= \boldsymbol{\Sigma}_i^{-1} \eta_i,\\
  \mu_j  &= \boldsymbol{\Sigma}_j^{-1} \eta_j,\\
  v_i    &= \frac{\alpha_i}{\alpha_i + \alpha_j},\\
  v_j    &= \frac{\alpha_j}{\alpha_i + \alpha_j}.
\end{align}
\end{subequations}
When multiple likelihoods share a common range-space, the next pair to be merged has the lowest upper bound of the relative Kullback--Leibler divergence given by
%
\begin{align}
   \bar{B}(i,j) =  (\alpha_i+\alpha_j)\ln |\boldsymbol{\Sigma}^{-1}_{ij}|  +\alpha_i\ln |\boldsymbol{\Sigma}_i| +\alpha_j\ln |\boldsymbol{\Sigma}_j|.
\end{align}
Using the above procedure, we can merge likelihood components within a
common range-space until a desired maximum number of components is
achieved. Following this reduction process, the generated components
can be transformed back into the original state-space using
\begin{subequations}
\label{eq:fjgheuf476}
\begin{align}
\mys &= \mathbf{U} \eta, \\
\myL &= \mathbf{U} \boldsymbol{\Sigma} \mathbf{U}^T,
\end{align}
\end{subequations}
where $\mathbf{U}$ is common to each component sharing the range-space.
This approach outlined by \eqref{eq:MCHA6100 Notes on JMLS:50}-\eqref{eq:fjgheuf476} is repeated for each range-space likelihood modes occupy.

\begin{figure}[!tbh]
\centering
\includegraphics[width=\linewidth]{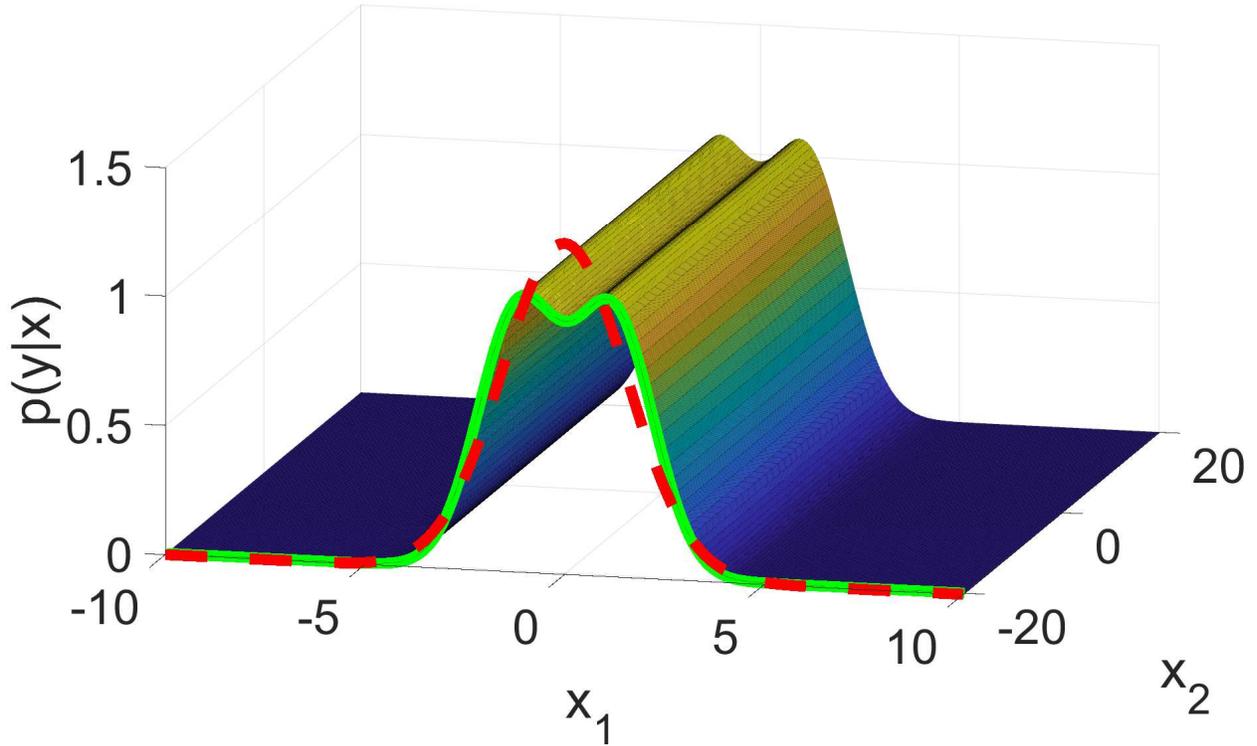}
\caption{A likelihood surface with two components comprising $p(y|x)$ is shown, which can also be described in a reduced space shown in solid green. After component reduction, the likelihood function is approximated in the reduced space by the dashed red line, before being transformed back into the full state-space.}
\label{fig:surfdemo1}
\end{figure}

For illustration, Figure~\ref{fig:surfdemo1} shows the scenario where two likelihood components have a common range-space. These likelihood components contain no information about state $x_2$ and have been superimposed to generate the surface mesh.
Using the method provided, a transformation was made to describe these likelihood components in terms of the common range-space before super imposing them to produce the solid green line.
Next the KLR method was used to combine these components in the range-space, resulting in the approximation shown in dotted red.
Following this, the approximated likelihood can be transformed back into the original 2D space.
Note that very seldom does the observed space align with the basis for the states, but this case is automatically handled by use of the SVD.
\subsection{Algorithm overview}
Algorithm~\ref{alg:JMLSsmoother} is provided to clarify the overall operation of the proposed solution to JMLS smoothing.
We recommend implementing the proposed algorithm using square-root factors and log-weights for numerically stability.
Additionally, for computational efficiency, components with zero weight do not need to be stored, and their statistics need not be calculated, this is often encountered when there is zero probability from transitioning from one model to another. 

\begin{algorithm}
\caption{The Two-filter JMLS Smoother}
\label{alg:JMLSsmoother}
\begin{algorithmic}[1]
\Require The JMLS system parameters $\{\mathbf{A}_k(z_k),\myB_k(z_k),\mathbf{Q}_k(z_k),\mathbf{C}_k(z_k),\myD_k(z_k),\mathbf{R}_k(z_k)\}_{z_k=1}^{m_k}$, defined model transition function $T_k$, measurement vector $y_k$ for timesteps $k=1,..,N$. The statistics for the prior $p(x_1,z_1)$ is also required. 
\For{$k=1,\dots,N$}
\State Calculate the statistics of the forward filtered distribution $p(x_k,z_k|y_{1:k})$ using Lemma~\ref{lem:fwdfilter45}.
\State Perform KL reduction, detailed in Subsection~\ref{sec:kullb-leibl-reduct}, over the subsets of components in $p(x_k,z_k|y_{1:k})$ with a common $z_k$ value.
\State Calculate the statistics of the prediction distribution $p(x_{k+1},z_{k+1}|y_{1:k})$ using Lemma~\ref{lem:fwdfilter45}.
\EndFor
\State Initialise the backwards information filter likelihood with $p(y_N|x_{N},z_N)$ using Lemma~\ref{lemma:backw-filt-smooth}.
\For{$k=N-1,\dots,1$}
\State Calculate the statistics of the backwards propagated likelihood $p(y_{k+1:N}|x_k,z_k)$ using Lemma~\ref{lemma:backw-filt-smooth}.
\State Using the method outlined in Subsection~\ref{sec:likelihood_red}, perform likelihood reduction over the subsets of likelihood components in $p(y_{k+1:N}|x_k,z_k)$ with a common range-space and $z_k$ value. 
\State Calculate the statistics of the corrected likelihood $p(y_{k:N}|x_k,z_k)$ using Lemma~\ref{lemma:backw-filt-smooth}.
\EndFor
\For{$k=1,\dots,N-1$}
\State Generate the statistics for the hybrid smoothed distribution $p(x_k,z_k|y_{1:N})$ using Lemma~\ref{lem:smoothing}.
\State (Optional) Perform KL reduction detailed in Subsection~\ref{sec:kullb-leibl-reduct}, over the subsets of components in $p(x_k,z_k|y_{1:N})$ with a common $z_k$ value.
\EndFor
%
\end{algorithmic}
\end{algorithm}

\clearpage
\section{Simulations}\label{sec:simulations}
Here we provide the results from smoothing three different systems to
demonstrate the effectiveness and versatility of the proposed
solution.
\subsection{Example 1 - Unimodal system}
\label{sec:example1}
In this example we consider the unimodal linear Gaussian system
\begin{subequations}
\begin{align}
x_{k+1} &= \mathbf{A}x_k + \myB u_k + v_k, \\
y_{k} &= \mathbf{C}x_k + \myD u_k + e_k, \\
  v_k &\sim \mathcal{N}(v_k \mid {0},\mathbf{Q}), \\
  e_k &\sim \mathcal{N}(e_k \mid {0},\mathbf{R}),
\end{align}
\end{subequations}
which can be thought of as a single model JMLS with a model transition
function of $T_k(z_{k+1} \mid z_k) =1$.
This system boycotts much of the difficulty with estimation of JMLS, as it does not require an exponentially growing number of components to be handled, and is included here for validation purposes of the smoothing equations.

\begin{figure}
\centering
\includegraphics[width=\linewidth]{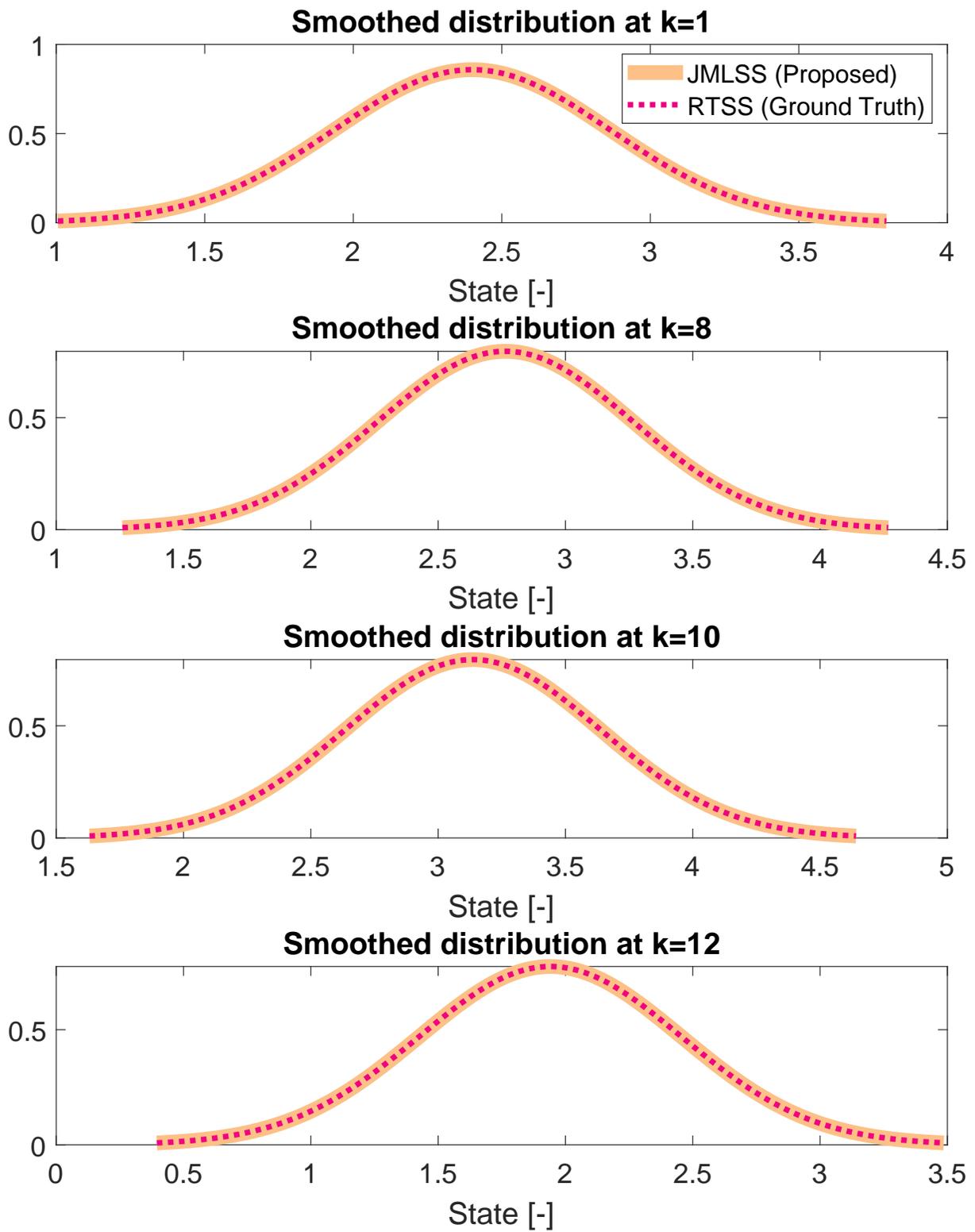}
\caption{Generated smoothed distribution for Example 1, where the Rauch--Tung--Striebel smoother (RTSS) density (dotted red) is used as the ground truth, which can be compared to the JMLS smoother (JMLSS) density (solid orange).}
\label{fig:Example1}
\end{figure}

The data for this example was generated using an input generated
according to ${u}_k\sim \mathcal{N}(0,1)$ and the parameters
\begin{align}
  \mathbf{A}&=0.9,&\, \myB&=0.1,&\, \mathbf{Q}&=0.45,\\
  \mathbf{C}&=0.9,&\, \myD&=0.05,&\, \mathbf{R}&=0.5.
\end{align}
Smoothing for this system was conducted for 13 timesteps using the proposed method and a RTS smoother to generate the ground truth data.
Resulting smoothed distributions from both of these methods is shown in Figure~\ref{fig:Example1} for comparison. As shown by this figure, both methods produce the same density, and are both exact for this system class.
\subsection{Example 2 - A Jump Markov linear system}
In this example we consider a JMLS system in the following form as used in \cite{helmick1995fixed,kim1994dynamic,doucet2001particle,barber2006expectation},
\begin{subequations}
\begin{align}
&x_{k} = \mathbf{A}_k(z_k)x_{k-1} + \myB_k(z_k) u_k + v_{k-1}, \\
&y_{k} = \mathbf{C}_k(z_k)x_k + \myD_k(z_k) u_k + e_k, \\
&v_{k-1} \sim \mathcal{N}(v_{k-1} \mid {0},\mathbf{Q}_k(z_k)), \\
&e_k \sim \mathcal{N}(e_k \mid {0},\mathbf{R}_k(z_k)), 
\end{align}
\end{subequations}
this form differs from the system described in \eqref{eq:JMLSdef1}, as
the above system uses a different time index for the switching
parameter in the prediction step.  To accommodate this, we modify the
forward filter and proposed backward filter such that the model
switches before the prediction step, and not after.

Since some of the alternate algorithms do not explain in detail how to
perform likelihood reduction in the case of non-integrable likelihood
functions, we use a single state system to circumvent this problem in
order to compare their performance to the proposed method.

The parameters used for data generation and smoothing of this system
were $u_k = 1$ for all $k$ and
\begin{align}
\mathbf{A}_k(1)&=0.9,\ \myB_k(1)=0.1,\ \mathbf{C}_k(1)=0.9,\ \myD_k(1)=0.05, \nonumber \\
\mathbf{A}_k(2)&=0.9,\ \myB_k(2)=0.12,\ \mathbf{C}_k(2)=0.85,\
              \myD_k(2)=0.05,\nonumber \\
\mathbf{Q}_k(1)&=0.45,\ \mathbf{R}_k(1)=0.5, \nonumber \\
\mathbf{Q}_k(2)&=0.01,\ \mathbf{R}_k(2)=1.5, \ 
\mathbf{T} = \begin{bmatrix}0.6 &0.4\\
              0.4 & 0.6\end{bmatrix},
\end{align}
where $T_k(i|j) = \mathbf{T}_{i,j}$.

\begin{figure}
\centering
\includegraphics[width=\linewidth]{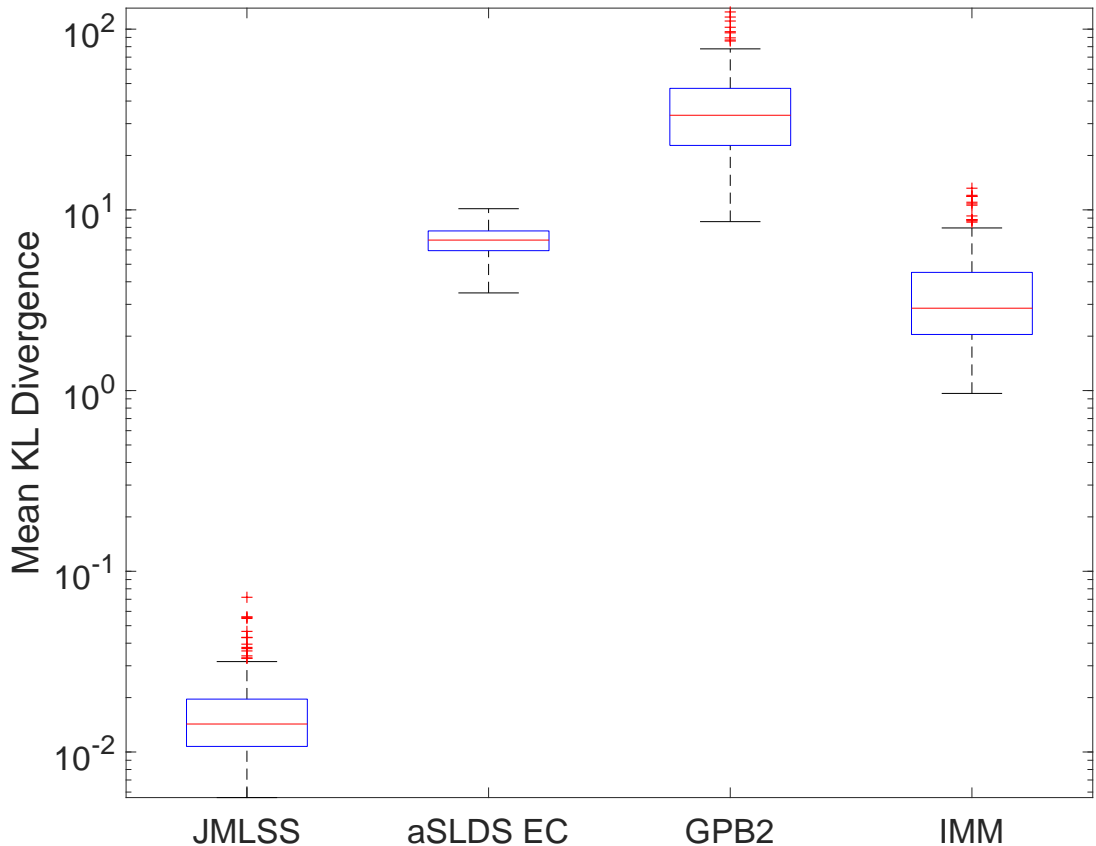}
\caption{Boxplot~\cite{mcgill1978tukey} of KL divergence over 250 random smoothed datasets for Example 2 using the proposed JMLS smoother (JMLSS), and other alternative IMM, GPB2, and aSLDS EC smoothers.}
\label{fig:boxplot2}
\end{figure}

\begin{figure*}
\centering
\includegraphics[width=\linewidth]{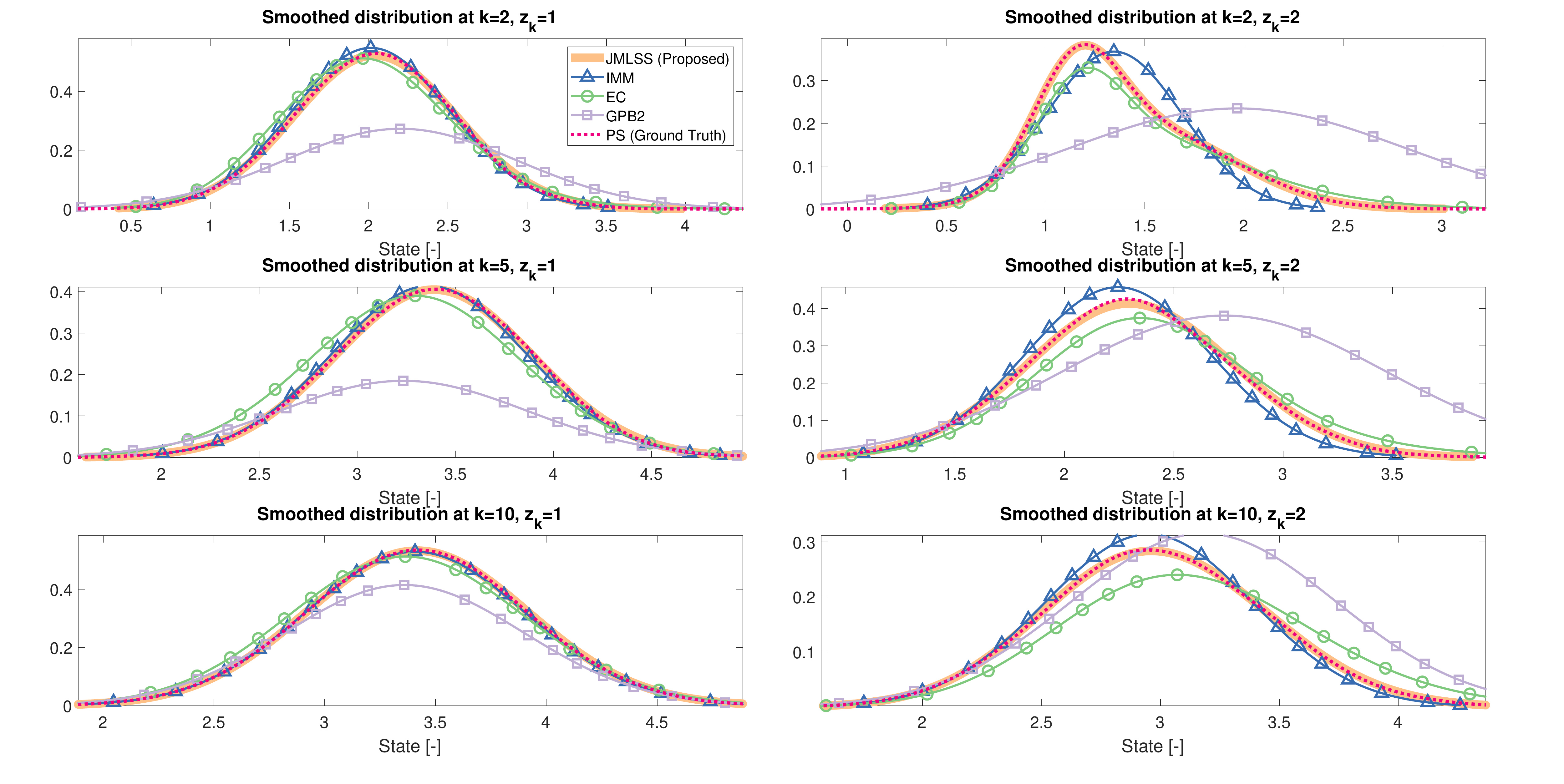}
\caption{Generated smoothed distributions for the JMLS system from Example 2, where the Particle Smoother (PS) density (dotted red) is used as the ground truth, and the distribution from the proposed JMLS smoother (JMLSS) (solid orange) is compared against the distributions from the alternate interacting multiple model (IMM) (blue with triangles), expectation correction (aSLDS EC) (green with circles), generalised pseudo-Bayesian (GPB2) (light purple with squares) smoothers.}
\label{fig:Example2}
\end{figure*}

The system was simulated for $N=15$ time steps to produce input-output
data. A range of smoothers were run on this data including the
proposed method (abbreviated as JMLSS) and other smothers including
the IMM, GPB2, and aSLDS EC smoother.  A particle smoother (PS)
solution was also implemented to provide a ground truth estimate,
however it took orders of magnitude longer to run.  This experiment
repeated 250 times with different datasets, where the distributions
were used to calculate a mean KL divergence error over each timeseries
for each of the methods, which in turn was used to generate the
boxplot in Figure~\ref{fig:boxplot2}.  Figure~\ref{fig:boxplot2} shows
that the proposed methods outperformed the alterative approaches in
each of these 250 runs, often by an order of magnitude.  Additionally
Figure~\ref{fig:Example2} shows distributions from each of the
smoothers during one of these runs.

It is important to also compare the computation time for these
methods, since it can be argued that the approximation error of the JMLS smoother
presented in this paper can always be reduced by allowing more
components. Table~\ref{tab:times} records the computation time for
each method (all methods were implemented in native Matlab code). It
may be concluded that the computation time associated with the JMLSS
is an order of magnitude more than the fastest method, but around two
orders of magnitude better in terms of accuracy.
\begin{table}[htb]
  \centering 
  \caption{\em Computation time (seconds) for Example~2.}
  {
  \begin{tabular}{c|c|c|c|c}
  	\toprule
    \textbf{PS} & \textbf{JMLSS} & \textbf{aSLSDS EC} & \textbf{GPB2}
    & \textbf{IMM}   \\
	\midrule
    16598 & 0.39 & 0.45 & 0.02 & 0.04\\
	\bottomrule
  \end{tabular}
}
  \label{tab:times}
\end{table}
\subsection{Example 3 - Dynamic JMLS system}
In this example we consider a two-state problem, which takes full
advantage of the proposed likelihood reduction strategy, as it is a
multi-state problem which presents with non-integrable likelihood
components in the BIF.  The example considers the dynamical model of a
mass-spring-damper (MSD) system with a position sensor
\begin{subequations}
\begin{align}
m\ddot{x}(t) + b\dot{x}(t) +kx(t) &= F(t), \\
y(t) &= x(t),
\end{align}
\end{subequations}
where $x(t)$ is the mass position, $F(t)$ is the applied external
force, $m$ is the system mass, $b$ is the damping coefficient and $k$
is the spring constant. We further assume that the system can use one
of two available sets of parameters, indicated by a superscript, at
any timestep,
\begin{align}m^1=8 [\textrm{kg}], \, b^1=12 [\textrm{Ns/m}], \, k^1=10 [\textrm{N/m}], \nonumber \\
m^2=8 [\textrm{kg}], \, b^2=0 [\textrm{Ns/m}], \, k^2=0 [\textrm{N/m}].
\end{align}
The second parameter set represents a possible fault scenario, where the the spring and damper have become disconnected from the mass, the transition matrix for this example was chosen to reflect a 1\% chance of permanent failure
\begin{align} \mathbf{T}_{i,j}=\begin{bmatrix}0.99 & 0 \\0.01 & 1 \end{bmatrix}.\end{align}
%
Using an approach similar to that in
\cite{ljung2010issues,wills2012estimation}, the models were
discretised and converted into the model class \eqref{eq:JMLSdef1}
using a sample rate of 100Hz and simulated ADC sample time of $0.1$ms.
The system was assumed to be driven by $F[k] = 2000\sin (k/(20\pi))$, then smoothed using the proposed method and a particle smoother (PS) to provide a ground truth.
As other alternative smoothers without modification do not support the system form \eqref{eq:JMLSdef1} required for this discretisation, they do not appear in this experiment. 

\begin{figure*}[p]
\centering
\includegraphics[width=\linewidth]{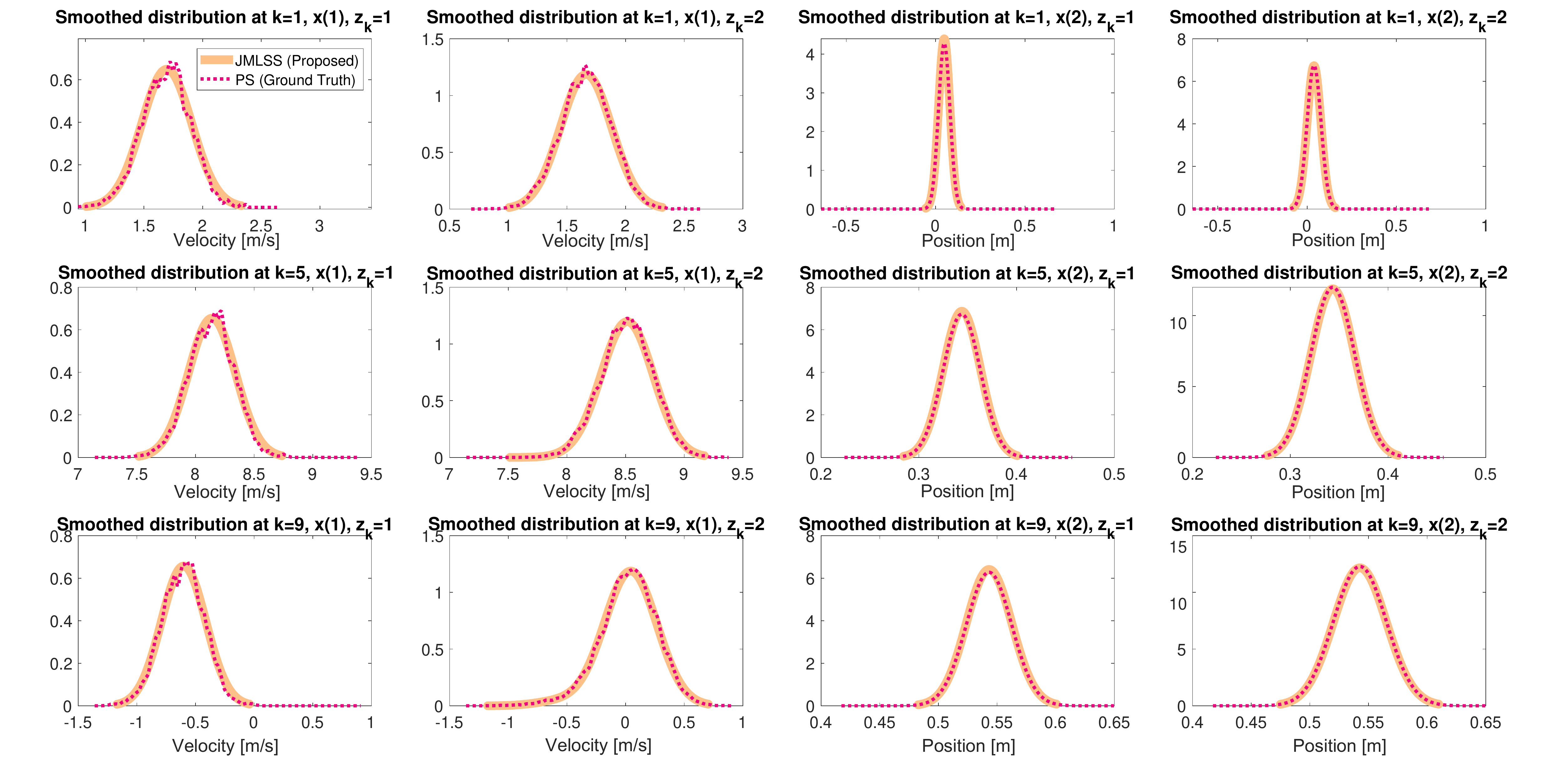}
\caption{Generated smoothed distributions for the dynamic JMLS system from Example 3, where the particle smoother (PS) distribution (dotted red) is used as the ground truth, and compared against the distribution from the proposed JMLS smoother (JMLSS) (solid orange).}
\label{fig:Example3}
\end{figure*}

The resulting marginalised distributions from this experiment are shown in Figure~\ref{fig:Example3}.
Note that the system could easily be smoothed for a larger number of timesteps using the proposed computationally inexpensive closed-form solution, but was set to $N=10$ due to the large computational expense of the particle smoother. It is also possible to increase number of models to accommodate a variety of other fault scenarios, such as sensor failure.
%
%
%
%
%
%
%
\clearpage
\section{Conclusion}\label{sec:conclusion}
We have developed a new smoothing algorithm for
jump-Markov-linear-systems using a two-filter approach.  This
contribution has two components.  Firstly, we have developed an
algorithm that implements the two-filter smoothing formulas
\textit{exactly}, therefore relaxing assumptions imposed by competing
methods.  Secondly, we have developed an approximation method that
allows likelihood reduction that, contrary to existing methods, does
not require likelihoods to have a Gaussian form in order to perform
merging. The implication is that the new method presents the user with
a mechanism to choose between computational cost and accuracy, making
it suitable for both real-time and post processing applications. This
accuracy-computational cost tradeoff was demonstrated by way of
example.
%

Compared to alternatives, the proposed approach is very well suited to
applications where likelihood modes in the BIF are non-integrable over
$x_k$. This can be encountered for a number of reasons, and is a
common occurrence in the first few iterations of the BIF. This
property allows the JMLS estimator to handle models with only partial
observability of the system state, which is useful for
applications in fault diagnosis.

\bibliographystyle{plain}        
\bibliography{autosam}           
\appendix
\clearpage
\section{Forward and backward filtering Lemmata}
\begin{small}
Recall the definition of $\xi \triangleq \{x_k, z_k\}$ from
Section~\ref{sec:problem-formulation}, and recall that for an integrable function $f$,
\begin{align}
  \label{eq:autosam:7}
  \int f(\xi_k) p(\xi_k) d\xi_k \triangleq \sum_{z_k=1}^{m_k} \int
  f(x_k,z_k) p(x_k,z_k) dx_k.
\end{align}
\subsection{Proof of Lemma~\ref{lemma:backw-filt-smooth}}
Begin by constructing
\begin{align}
p&(y_{k+1:N}|\xi_k) \nonumber \\
=& \int p(y_{k+1:N}|\xi_{k+1})p(\xi_{k+1}|\xi_k) \, d\xi_{k+1} \nonumber \\
=& \int \sum_{i=1}^\numCorr{k+1} \mathcal{L}(x_{k+1}| \bar{r}^i_{k+1}(z_{k+1}),\bar{s}^i_{k+1}(z_{k+1}),\bar{\myL}^i_{k+1}(z_{k+1})) \nonumber \\
&\cdot T(z_{k+1}|z_k) \mathcal{N}(x_{k+1}|\mathbf{A}_k(z_k)x_k+b_k(z_k),\mathbf{Q}_k(z_k)) \, d\xi_{k+1} \nonumber \\
=& \sum_{z_{k+1}=1}^{m_{k+1}} \int \sum_{i=1}^\numCorr{k+1} \mathcal{L}(x_{k+1}| \bar{r}^i_{k+1}(z_{k+1}),\bar{s}^i_{k+1}(z_{k+1}),\bar{\myL}^i_{k+1}(z_{k+1})) \nonumber \\
&\cdot T(z_{k+1}|z_k) \mathcal{N}(x_{k+1}|\mathbf{A}_k(z_k)x_k+b_k(z_k),\mathbf{Q}_k(z_k)) \, d\xi_{k+1} \nonumber \\
=& \sum_{z_{k+1}=1}^{m_{k+1}} \sum_{i=1}^\numCorr{k+1} T(z_{k+1}|z_k) \nonumber \\
&\cdot \int\mathcal{L}(x_{k+1}| \bar{r}^i_{k+1}(z_{k+1}),\bar{s}^i_{k+1}(z_{k+1}),\bar{\myL}^i_{k+1}(z_{k+1})) \nonumber \\
&\cdot  \mathcal{N}(x_{k+1}|\mathbf{A}_k(z_k)x_k+b_k(z_k),\mathbf{Q}_k(z_k)) \, d\xi_{k+1}.
\end{align}
Applying Lemma~\ref{lem:backpropalemma} yields
\begin{align}
=& \sum_{z_{k+1}=1}^{m_{k+1}} \sum_{i=1}^\numCorr{k+1} T(z_{k+1}|z_k) \nonumber \\
& \cdot \mathcal{L}(x_{k}| \tilde{r}^i_{k}(z_k,z_{k+1}),{s}^i_{k}(z_k,z_{k+1}),\myL^i_{k}(z_k,z_{k+1}))  .
\end{align}
Absorbing the transition probability into the information scalar yields
\begin{align}
=& \sum_{z_{k+1}=1}^{m_{k+1}} \sum_{i=1}^\numCorr{k+1} \mathcal{L}(x_{k}| {r}^i_{k}(z_k,z_{k+1}),{s}^i_{k}(z_k,z_{k+1}),\myL^i_{k}(z_k,z_{k+1}))  ,
\end{align}
where
\begin{align}
{r}^i_{k}(z_k,z_{k+1}) = \tilde{r}^i_{k}(z_k,z_{k+1}) -2\ln(T(z_{k+1}|z_k)).
\end{align}
Finally, the double sum indices $\{z_{k+1},i\}$ can be collapsed into single index $j$,
\begin{align}
=& \sum_{j=1}^\numBack{k} \mathcal{L}(x_{k}| r^j_k(z_k),s^j_k(z_k),\myL^j_k(z_k))  ,
\end{align}
where $\numBack{k}=m_{k+1}\numCorr{k+1}$.
%
\\\\
The proof for the expressions in \eqref{label:likerecursion} and \eqref{eq:MCHA6100 Notes on JMLS:201} is provided below
\begin{align}
p&(y_{k+1:N}|\xi_{k+1}) \nonumber\\
=& p(y_{k+2:N}|\xi_{k+1})p(y_{k+1}|\xi_{k+1}) \nonumber\\
=& \sum_{i=1}^{\numBack{k+1}} \mathcal{L}(x_{k+1}|r_{k+1}^i(z_{k+1}), s_{k+1}^i(z_{k+1}),\myL_{k+1}^i(z_{k+1})) \nonumber \\
& \cdot \mathcal{N}(y_{k+1}|\mathbf{C}_{k+1}(z_{k+1})x_{k+1} +d_{k+1}(z_{k+1}), \mathbf{R}_{k+1}(z_{k+1})).
\end{align}
Using Lemma~\ref{lem:meascorrlem},
\begin{align}
=& \sum_{i=1}^{\numBack{k+1}} \mathcal{L}(x_{k+1}|\bar{r}_{k+1}^i(z_{k+1}), \bar{s}_{k+1}^i(z_{k+1}),\bar{\myL}_{k+1}^i(z_{k+1})),
\end{align}
finally let $\numCorr{k+1}=\numBack{k+1}$,
\begin{align}
=& \sum_{i=1}^{\numCorr{k+1}} \mathcal{L}(x_{k+1}|\bar{r}_{k+1}^i(z_{k+1}), \bar{s}_{k+1}^i(z_{k+1}),\bar{\myL}_{k+1}^i(z_{k+1})).
\end{align}
\qed
\end{small}
\subsection{Proof of Lemma~\ref{lem:smoothing}}
\begin{small}
Begin by constructing 
 \begin{align}
 p&(y_{k+1:N}|\xi_k)p(\xi_k|y_{1:k}) \nonumber \\
=&\sum_{\ell=1}^{\numBack{k}} \mathcal{L}(x_{k}|\bar{r}_{k}^\ell(z_{k}), \bar{s}_{k}^\ell(z_{k}),\bar{\myL}_{k}^\ell(z_{k}))  \nonumber \\
 &\cdot \sum_{i=1}^\numFilt{k} w_{k|k}^i(z_k) \mathcal{N}(x_k|\mu_{k|k}^i(z_k),\myP_{k|k}^i(z_k)),
\end{align}
before using Lemma~\ref{lem:smoothedmode} to yield
 \begin{align}
=&\sum_{\ell=1}^{\numBack{k}} \sum_{i=1}^\numFilt{k} \tilde{w}_{k|N}^{(i,\ell)}(z_k) \mathcal{N}(x_k|\mu_{k|N}^{(i,\ell)}(z_k),\myP_{k|N}^{(i,\ell)}(z_k)).
\end{align}
Collapse double sum indices $\{ i,\ell \}$ into single sum $j$
 \begin{align}
\label{eq:eqsmoothedcomp11}
=&\sum_{j=1}^{\numSmooth{k}} \tilde{w}_{k|N}^{j}(z_k) \mathcal{N}(x_k|\mu_{k|N}^{j}(z_k),\myP_{k|N}^{j}(z_k)),
\end{align}
where $\numSmooth{k} = \numFilt{k} \numBack{k}$. Now concentrating on the smoothed distribution
\begin{align}
 p(\xi_k|y_{1:N})  =&\frac{ p(y_{k+1:N}|\xi_k)p(\xi_k|y_{1:k})}{p(y_{k+1:N}|y_{1:k})} \nonumber \\
=&\frac{ p(y_{k+1:N}|\xi_k)p(\xi_k|y_{1:k})}{\int p(y_{k+1:N}|\xi_k)p(\xi_k|y_{1:k}) \, d\xi_k } \nonumber \\
=&\frac{ p(y_{k+1:N}|\xi_k)p(\xi_k|y_{1:k})}{\sum_{z_k=1}^{m_k}\int p(y_{k+1:N}|\xi_k)p(\xi_k|y_{1:k}) \, dx_k } ,
 \end{align}
now substituting \eqref{eq:eqsmoothedcomp11} yields
\begin{align}
=&\frac{ \sum_{j=1}^{\numSmooth{k}} \tilde{w}_{k|N}^{j}(z_k) \mathcal{N}(x_k|\mu_{k|N}^{j}(z_k),\myP_{k|N}^{j}(z_k)) }{\sum_{z_k=1}^{m_k}\int \sum_{j=1}^{\numSmooth{k}} \tilde{w}_{k|N}^{j}(z_k) \mathcal{N}(x_k|\mu_{k|N}^{j}(z_k),\myP_{k|N}^{j}(z_k)) \, dx_k }  \nonumber \\
=&\frac{ \sum_{j=1}^{\numSmooth{k}} \tilde{w}_{k|N}^{j}(z_k) \mathcal{N}(x_k|\mu_{k|N}^{j}(z_k),\myP_{k|N}^{j}(z_k)) }{\sum_{z_k=1}^{m_k} \sum_{j=1}^{\numSmooth{k}} \tilde{w}_{k|N}^{j}(z_k) } \nonumber \\
=&  \sum_{j=1}^{\numSmooth{k}} {w}_{k|N}^{j}(z_k) \mathcal{N}(x_k|\mu_{k|N}^{j}(z_k),\myP_{k|N}^{j}(z_k)) ,
\end{align}
where 
\begin{align}
{w}_{k|N}^{j}(z_k) = \frac{ \tilde{w}_{k|N}^{j}(z_k)  }{\sum_{z_k=1}^{m_k} \sum_{j=1}^{\numSmooth{k}} \tilde{w}_{k|N}^{j}(z_k) }.
\end{align}
\qed
\end{small}
\clearpage
\section{Additional Lemata }
\begin{lem}
\label{lem:backpropalemma}
\begin{small}
Let ${\bar{\mathbf{L}}}$, $\bar{s}$, and $\bar{r}$ be the information matrix, information vector and information scalar respectively, the sufficient statistics for a likelihood component in the BIF.
Then given a Gaussian state-transition distribution parameterised by $\mathbf{A}$, and $b$, with an invertible process covariance $\mathbf{Q}$,
then the statistics for the BIF likelihood can be backwards propagated as
\begin{align}&\mathcal{L}(x|r,s,\myL)  \nonumber \\
&=\int  \mathcal{N}(\myinte|\mathbf{A}x+b,\mathbf{Q}) \mathcal{L}(\myinte|\bar{r},\bar{s},\bar{\myL}) \, d\myinte,\end{align}
where ${\mathbf{L}}$, ${s}$, and ${r}$  are the updated statistics. These statistics can be calculated using
\begin{align}
    \myL&=\myA^T \myPhi \myA,\\
    s& = \myA^T(\myPhi b +\myBeta^T\bar{s} ) ,\\ 
       r& = \bar{r} - \ln{|\myBeta|}   + \begin{bmatrix}\bar{s} \\ b\end{bmatrix}^T \begin{bmatrix}\myPsi & \myBeta \\    \myBeta^T & \myPhi \end{bmatrix} \begin{bmatrix}\bar{s} \\ b\end{bmatrix},\\
    \myBeta& = \myI- \myQ \myPhi, \\
    \myPsi& = \myQ \myPhi \myQ - \myQ,\\
    \myPhi&=\Bigl (\myI + \bar{\myL} \myQ \Bigr )^{-1} \bar{\myL}.
\end{align}
\end{small}
\end{lem}
\begin{pf}
\begin{small}
Begin with the Gaussian state-transition distribution
\begin{align}\mathcal{N}&(\myinte|\mathbf{A}x+b,\mathbf{Q}) \nonumber \\
=& e^{-0.5(x^p - \myinte)^T\mathbf{Q}^{-1}(x^p - \myinte)-0.5\ln |2\pi \mathbf{Q}|} \nonumber \\
=&e^{-0.5(   \myinte^T\mathbf{Q}^{-1}\myinte - 2\myinte^T\mathbf{Q}^{-1}x^p + (x^p)^T\myQ^{-1}x^p +\ln |2\pi \myQ|)},\end{align}
where $x^p=\myA x+b.$
Therefore
\begin{align}
\int & \mathcal{N}(\myinte |\myA x+b,\myQ) \mathcal{L}(\myinte|\bar{r},\bar{s},\bar{\myL}) \, d\myinte \nonumber \\
&=\int \mathcal{N}(\myinte |\myA x+b,\myQ) e^{-0.5(\myinte^T\bar{\myL}\myinte+2\myinte^T\bar{s}+\bar{r})} \, d\myinte \nonumber \\
&=\int e^{-0.5(\myinte^T\hat{\myL}\myinte+2\myinte^T\hat{s}+\hat{r})} \, d\myinte \nonumber \\
&=\int e^{-0.5\hat{r}}e^{-0.5(\myinte^T\hat{\myL}\myinte+2\myinte^T\hat{s})} \, d\myinte \label{eq:intmepls4},
\end{align}
where,
\begin{subequations}
\label{eq:subsbackforbrevity}
\begin{align}\hat{\myL} &= \bar{\myL} + \myQ^{-1} ,\\
\hat{s}&= \bar{s} - \myQ^{-1}x^p,\label{eq:shat}\\
\hat{r}&= \bar{r} + (x^p)^T\myQ^{-1}x^p +\ln |2\pi \myQ|\label{eq:rhat}.
\end{align}
\end{subequations}
%
%
Using the normalising constant for a Gaussian we can derive an expression to perform the required integration,
\begin{align}|2\pi \myP|^{0.5} = \int e^{-0.5(x^T\myP^{-1}x-2x^T\myP^{-1}\mu+\mu^T\myP^{-1}\mu)} \, dx,\end{align} 
\begin{align}\therefore \int e^{-0.5(x^T\myP^{-1}x-2x^T\myP^{-1}\mu)}   \, dx = |2\pi \myP|^{0.5}e^{0.5\mu^T\myP^{-1}\mu}.\end{align}
Substituting in the relationship to information form, \\$\myP^{-1} = \hat{\myL}$, $\mu=-\hat{\myL}^{-1}\hat{s}$, and noting that $\hat{\myL}^T=\hat{\myL}$ yields
\begin{align}\int e^{-0.5(x^T\hat{\myL}x+2x^T\hat{\myL}\hat{\myL}^{-1}\hat{s})}   \, dx = |2\pi \hat{\myL}^{-1}|^{0.5}e^{0.5\hat{s}^T\hat{\myL}^{-1}\hat{\myL}\hat{\myL}^{-1}\hat{s}},\end{align}
\begin{align}\int e^{-0.5(x^T\hat{\myL}x+2x^T\hat{s})}   \, dx = |2\pi \hat{\myL}^{-1}|^{0.5}e^{0.5\hat{s}^T\hat{\myL}^{-1}\hat{s}}.\end{align}
We can now use this result to integrate \eqref{eq:intmepls4} over $\myinte$,
\begin{align}\int  &\mathcal{N}(\myinte|\myA x+b,\myQ)  \mathcal{L}(\myinte|\bar{r},\bar{s},\bar{\myL}) \, d\myinte\nonumber\\
&= e^{-0.5\hat{r}} \int e^{-0.5(\myinte^T\hat{\myL}\myinte+2\myinte^T\hat{s})} \, d\myinte\nonumber\\
&=e^{-0.5\hat{r}}|2\pi \hat{\myL}^{-1}|^{0.5}e^{0.5\hat{s}^T\hat{\myL}^{-1}\hat{s}}\nonumber\\
&=e^{-0.5(\hat{r}-\hat{s}^T\hat{\myL}^{-1}\hat{s}-\ln|2\pi \hat{\myL}^{-1}|)}.
\end{align}
%
Back substituting \eqref{eq:shat} and \eqref{eq:rhat}
\begin{align}\label{eq:last347}
=&\exp \bigg[-0.5 \bigg( \bar{r} + (x^p)^T\myQ^{-1}x^p +\ln |2\pi \myQ| \nonumber \\
&-(\bar{s} - \myQ^{-1}x^p)^T\hat{\myL}^{-1}(\bar{s} - \myQ^{-1}x^p)-\ln|2\pi \hat{\myL}^{-1}|   \bigg)\bigg] \nonumber
\\
=&\exp \bigg[-0.5 \bigg( \bar{r} +\ln |2\pi \myQ| -\ln|2\pi \hat{\myL}^{-1}| + (x^p)^T\myQ^{-1}x^p  \nonumber \\
& -\bar{s}^T\hat{\myL}^{-1}\bar{s}    +  2 \bar{s}^T\hat{\myL}^{-1}\myQ^{-1}x^p  - (x^p)^T\myQ^{-1}\hat{\myL}^{-1} \myQ^{-1}x^p   \bigg)\bigg] \nonumber
\\
=&\exp \bigg[-0.5 \bigg( \bar{r} +\ln |2\pi \myQ| -\ln|2\pi \hat{\myL}^{-1}| -\bar{s}^T\hat{\myL}^{-1}\bar{s}  \nonumber \\
&  + (x^p)^T(\myQ^{-1}-\myQ^{-1}\hat{\myL}^{-1} \myQ^{-1})x^p    +  2 \bar{s}^T\hat{\myL}^{-1}\myQ^{-1}x^p    \bigg)\bigg] \nonumber
\\
=&\exp \bigg[-0.5 \bigg( \bar{r} +\ln | \myQ| -\ln| \hat{\myL}^{-1}| -\bar{s}^T\hat{\myL}^{-1}\bar{s}  \nonumber \\
&  + (x^p)^T(\myQ^{-1}-\myQ^{-1}\hat{\myL}^{-1} \myQ^{-1})x^p    +  2 \bar{s}^T\hat{\myL}^{-1}\myQ^{-1}x^p    \bigg)\bigg].
\end{align}
Using the Woodbury matrix identity: \\$A^{-1}-(UCV+A)^{-1}=A^{-1}U(C^{-1}+VA^{-1}U)^{-1}VA^{-1}$
with $U=C=\mathbf{I}$, $V=\bar{\myL}$, and $A=\mathbf{Q}^{-1}$ yields the useful relations
\begin{align}\mathbf{Q} -(\underbrace{\bar{\myL} + \mathbf{Q}^{-1}}_{\hat{\myL}})^{-1} = \myQ\underbrace{(\mathbf{I}+\bar{\myL}\myQ)^{-1}\bar{\myL}}_{\myPhi}\myQ \label{eq:eqre3},\end{align}
and 
\begin{align}\label{eq:last34876}\therefore \mathbf{Q}^{-1} -\mathbf{Q}^{-1}(\underbrace{\bar{\myL}+\mathbf{Q}^{-1}}_{\hat{\myL}})^{-1}\mathbf{Q}^{-1} = (\mathbf{I}+\bar{\myL}\myQ)^{-1}\bar{\myL} = \myPhi.\end{align}
Substituting \eqref{eq:last34876} into \eqref{eq:last347}
\begin{align} \label{eq:backtothefuture}
=&\exp \bigg[-0.5 \bigg( \bar{r} +\ln | \myQ| -\ln| \hat{\myL}^{-1}| -\bar{s}^T\hat{\myL}^{-1}\bar{s}  \nonumber \\
&  + (x^p)^T\myPhi x^p    +  2\bar{s}^T\hat{\myL}^{-1}\myQ^{-1}x^p    \bigg)\bigg].
\end{align}
%
%
From \eqref{eq:eqre3} $\myQ-\hat{\myL}^{-1} = \myQ \myPhi \myQ $, $\therefore\hat{\myL}=(\myQ - \myQ\myPhi \myQ)^{-1}$, and therefore
\begin{align}&\ln | \myQ| -\ln| \hat{\myL}^{-1}| =\ln | \myQ \hat{\myL}|\nonumber \\
&=\ln|\myQ(\myQ - \myQ\myPhi \myQ)^{-1}|=\ln|(\mathbf{I} - \myQ\myPhi )^{-1}| \nonumber \\
&=-\ln |\myBeta|, \text{where } \myBeta = (\mathbf{I} - \myQ\myPhi ),
 \end{align}
which can be substituted into \eqref{eq:backtothefuture} to yield
\begin{align} \label{eq:oldderive3443}
=&\exp \bigg[-0.5 \bigg( \bar{r} -\ln| \myBeta| -\bar{s}^T\hat{\myL}^{-1}\bar{s}  + (x^p)^T\myPhi x^p    \nonumber \\
&    +  2\bar{s}^T\hat{\myL}^{-1}\myQ^{-1}x^p    \bigg)\bigg].
\end{align}
By rearrangement of \eqref{eq:eqre3}
\begin{align}\hat{\myL}^{-1}\myQ^{-1} = ( \myQ - \myQ\myPhi \myQ) \myQ^{-1} =\mathbf{I} - \myQ\myPhi =\myBeta,\label{eq:backeer1} \end{align}
and
\begin{align}\hat{\myL}^{-1} = \myQ - \myQ\myPhi \myQ = -\myPsi \label{eq:backeer}.\end{align}
Substituting \eqref{eq:backeer1} and \eqref{eq:backeer} into \eqref{eq:oldderive3443} yields
\begin{align} 
=&\exp \bigg[-0.5 \bigg( \bar{r} -\ln| \myBeta| +\bar{s}^T\myPsi\bar{s}  
&  + (x^p)^T\myPhi x^p    +  2\bar{s}^T\myBeta x^p    \bigg)\bigg].
\end{align}
Finally, by back substituting $x^p=\myA x+b$, 
\begin{align} 
=&\exp \bigg[-0.5 \bigg( \bar{r} -\ln| \myBeta| +\bar{s}^T\myPsi\bar{s}  \nonumber \\
&  + (\myA x+b)^T\myPhi (\myA x+b)   +  2\bar{s}^T\myBeta (\myA x+b)    \bigg)\bigg]\\
=&\exp \bigg[-0.5 \bigg( \bar{r} -\ln| \myBeta| +\bar{s}^T\myPsi\bar{s}  \nonumber \\
&  + x^T\myA^T\myPhi\myA x + 2b^T\myPhi\myA x +b^T\myPhi b   +  2\bar{s}^T\myBeta \myA x+  2\bar{s}^T\myBeta b    \bigg)\bigg]\\
=&\exp \bigg[-0.5 \bigg( \bar{r} -\ln| \myBeta| +\bar{s}^T\myPsi\bar{s} +  \bar{s}^T\myBeta b +b^T \myBeta^T\bar{s} +b^T\myPhi b \nonumber \\
&  + x^T\myA^T\myPhi\myA x + 2(b^T\myPhi\myA +\bar{s}^T\myBeta \myA)x        \bigg)\bigg]\\
=&\exp \bigg[-0.5 \bigg( x^T\underbrace{\myA^T\myPhi\myA}_\myL x + 2x^T\underbrace{\myA^T(\myPhi b +\myBeta^T\bar{s} )   }_s \nonumber \\
&     + \underbrace{\bar{r} -\ln| \myBeta| + \begin{bmatrix}\bar{s} \\ b\end{bmatrix}^T \begin{bmatrix}\myPsi & \myBeta \\    \myBeta^T & \myPhi \end{bmatrix} \begin{bmatrix}\bar{s} \\ b\end{bmatrix}    }_r \bigg)\bigg], \nonumber \\
=& \mathcal{L}(x|r,s,\myL).
\end{align}
Note that unlike $\myPsi$ and $\myPhi$, $\myBeta$ is \emph{not} necessarily symmetric.
\qed
\end{small}
\end{pf}
\begin{lem}
\label{lem:meascorrlem}
\begin{small}
Let ${\mathbf{L}}$, ${s}$, and ${r}$ be the information matrix, information vector and information scalar respectively, the sufficient statistics for a likelihood component in the BIF.
Then given a measurement $y$, and the parameters for the output equation $\mathbf{C}$, and $d$, with an invertible measurement covariance $\mathbf{R}$,
then the statistics for the BIF likelihood component can be corrected by the a Gaussian likelihood as 
\begin{align} \mathcal{L}(x|\bar{r},\bar{s},\bar{\mathbf{L}}) = \mathcal{N}(y|\mathbf{C}x+d,\mathbf{R}) \mathcal{L}(x|{r},{s},{\mathbf{L}}) ,\end{align}
where $\bar{\mathbf{L}}$, $\bar{s}$, and $\bar{r}$  are the corrected statistics. These statistics can be calculated using
\begin{align}
\bar{\mathbf{L}} &= \mathbf{L}  + \mathbf{C}^T\mathbf{R}^{-1}\mathbf{C},\\
\bar{s} &= s + \mathbf{C}^T\mathbf{R}^{-1}\zeta,\\
\bar{r} &= r + \zeta^T\mathbf{R}^{-1}\zeta + \ln |2\pi \mathbf{R}|,\\
\zeta&=d-y.
\end{align}
\end{small}
\end{lem}
\begin{pf}
\begin{small}
Begin with the linear Gaussian measurement likelihood
\begin{align}
\mathcal{N}(&y|\mathbf{C}x+d,\mathbf{R}) \nonumber \\
=& e^{-0.5(x^T\mathbf{C}^T+\zeta^T)\mathbf{R}^{-1}(\mathbf{C}x+\zeta)-0.5\ln |2\pi \mathbf{R}|} \nonumber \\
=&\exp [-0.5(x^T\mathbf{C}^T\mathbf{R}^{-1}\mathbf{C}x+2x^T\mathbf{C}^T\mathbf{R}^{-1}\zeta \nonumber \\
&+\zeta^T\mathbf{R}^{-1}\zeta+ \ln |2\pi \mathbf{R}|)].
\end{align}
Therefore
\begin{align}
\mathcal{N}&(y|\mathbf{C}x+d,\mathbf{R})\mathcal{L}(x|{r},{s},{\mathbf{L}})\nonumber \\
=&\mathcal{N}(y|\mathbf{C}x+d,\mathbf{R})e^{-0.5(x^T\mathbf{L}x+2x^Ts+r)} \nonumber \\
=&\exp [-0.5(x^T(\mathbf{L}+\mathbf{C}^T\mathbf{R}^{-1}\mathbf{C})x+2x^T(s+\mathbf{C}^T\mathbf{R}^{-1}\zeta) \nonumber \\
&+r+\zeta^T\mathbf{R}^{-1}\zeta+ \ln |2\pi \mathbf{R}|)]  \nonumber \\
=&e^{-0.5(x^T\bar{\mathbf{L}}x+2x^T\bar{s}+\bar{r})}=\mathcal{L}(x|\bar{r},\bar{s},\bar{\mathbf{L}}).
\end{align}
\qed
\end{small}
\end{pf}
\begin{lem}
\label{lem:smoothedmode}
Let ${\bar{\mathbf{L}}}$, $\bar{s}$, and $\bar{r}$ be the information matrix, information vector and information scalar respectively, the sufficient statistics for a likelihood component in the BIF.
Additionally, let $\myP$, $\mu$, and $w$ be the covariance matrix, mean and weight of a Gaussian mode. Then the statistics of the combined smoothed component $\{\bar{\myP}, \bar{\mu}, \bar{w} \}$ can be computed as
\begin{align} \bar{w} \mathcal{N}(x|\bar{\mu},\bar{\myP}) = w\mathcal{N} (x|\mu,\myP) \mathcal{L}(x|r,s,\myL),\end{align}
where
\begin{subequations}
\begin{align}
\bar{\myP} &= (\myP^{-1} + \myL)^{-1},\\
\bar{\mu} &= \bar{\myP}(\myP^{-1}\mu - s),\\
\bar{w} &= \frac{w{|2\pi \bar{\myP}|}^{\frac{1}{2}}e^{\frac{1}{2}\beta}}{{|2\pi \myP|}^{\frac{1}{2}}}, \\
\beta &= \bar{\mu}^T\bar{\myP}^{-1}\bar{\mu} - \mu^T \myP^{-1}\mu - r
\end{align}
\end{subequations}
\end{lem}
\begin{pf}
\begin{small}
\begin{align}
w&\mathcal{N} (x|\mu,\myP) \mathcal{L}(x|r,s,\myL) \nonumber \\
&=\frac{w}{{|2\pi \myP|}^{\frac{1}{2}}}e^{-\frac{1}{2} (x^T\myP^{-1}x-2x^T\myP^{-1}\mu + \mu^T\myP^{-1}\mu)}e^{-\frac{1}{2}(x^T\myL x+2x^Ts + r)} \nonumber \\
&=\frac{we^{-\frac{1}{2}(\mu^T\myP^{-1}\mu + r)}}{{|2\pi \myP|}^{\frac{1}{2}}}e^{-\frac{1}{2}\left(x^T(\myP^{-1}+\myL)x-2x^T(\myP^{-1}\mu-s)\right)} 
\end{align}
let $\bar{\myP}^{-1} = \myP^{-1} + \myL$, and $\bar{\mu} = \bar{\myP}(\myP^{-1}\mu - s)$
\begin{align}
=&\frac{we^{-\frac{1}{2}(\mu^T\myP^{-1}\mu + r)}}{{|2\pi \myP|}^{\frac{1}{2}}} e^{-\frac{1}{2}(x^T\bar{\myP}^{-1}x-2x^T\bar{\myP}^{-1}\bar{\mu})} \nonumber \\
=&\frac{w{|2\pi \bar{\myP}|}^{\frac{1}{2}}e^{-\frac{1}{2}(\mu^T\myP^{-1}\mu + r -\bar{\mu}^T\bar{\myP}^{-1}\bar{\mu})}}{{|2\pi \myP|}^{\frac{1}{2}}{|2\pi \bar{\myP}|}^{\frac{1}{2}}} \nonumber \\
&\cdot e^{-\frac{1}{2}(x^T\bar{\myP}^{-1}x-2x^T\bar{\myP}^{-1}\bar{\mu} +\bar{\mu}^T\bar{\myP}^{-1}\bar{\mu})}
\end{align}
let $\beta = \bar{\mu}^T\bar{\myP}^{-1}\bar{\mu} - \mu^T \myP^{-1}\mu - r$, and substitute for $\mathcal{N}(x|\bar{\mu},\bar{\myP})$
\begin{align}
&=\frac{w{|2\pi \bar{\myP}|}^{\frac{1}{2}}e^{\frac{1}{2}\beta}}{{|2\pi \myP|}^{\frac{1}{2}}} \mathcal{N}(x|\bar{\mu},\bar{\myP})
\end{align}
\qed
\end{small}
\end{pf}

\end{document}